\documentclass[fleqn,usenatbib]{mnras}

\usepackage{newtxtext,newtxmath}

\usepackage[T1]{fontenc}

\DeclareRobustCommand{\VAN}[3]{#2}
\let\VANthebibliography\thebibliography
\def\thebibliography{\DeclareRobustCommand{\VAN}[3]{##3}\VANthebibliography}

\usepackage{graphicx}	
\usepackage{amsmath}	
\usepackage{lineno}
\usepackage{natbib}

\title[EDGES and JWST with 21cm signal]{EDGES and JWST with 21cm global signal emulator} 

\author[S. Yoshiura et al.]{
Shintaro Yoshiura,$^{1}$\thanks{E-mail: shintaro.yoshiura@nao.ac.jp}\thanks{JSPS Research Fellow}
Teppei Minoda$^{2}$
and Tomo Takahashi$^{3}$
\\
$^{1}$Mizusawa VLBI Observatory, National Astronomical Observatory Japan, 2-21-1 Osawa, Mitaka, Tokyo 181-8588, Japan\\
$^{2}$The University of Melbourne, School of Physics, Parkville, VIC 3010, Australia\\
$^{3}$Department of Physics, Saga University, Saga 840-8502, Japan
}

\date{Accepted XXX. Received YYY; in original form ZZZ}

\pubyear{2015}

\begin{document}
\label{firstpage}
\pagerange{\pageref{firstpage}--\pageref{lastpage}}
\maketitle

\begin{abstract}

The 21cm global signal is an important probe to reveal the properties of the first astrophysical objects and the processes of the structure formation from which one can constrain astrophysical and cosmological parameters. To extract the information of such parameters, one needs to efficiently evaluate the 21cm global signal for statistical analysis. First we developed an artificial neural network-based emulator to predict the 21cm global signal, which works 
with significantly less computational cost and high precision. Then we apply our emulator to demonstrate the parameter estimation based on the Bayesian analysis by using the publicly available EDGES low-band data. We find that the result is sensitive to the foreground model, the assumption of noise, and the frequency range used in the analysis. The Bayesian evidence suggests the models with higher order polynomial function and enhanced noise are preferred. We also compare models suggested from the EDGES low-band data and the ones from recent JWST measurements of the galaxy luminosity function at $z=16$. We find that the model which produces the 21cm absorption line at $z\approx15$ is well consistent with the central value of the observed luminosity function at $z=16$.  

\end{abstract}




\section{Introduction}

During the Cosmic Dawn, when the first stars and galaxies are formed, the intergalactic medium (IGM) was filled with the neutral hydrogen atom. Thus, the 21cm line, emitted from the neutral hydrogen atoms, can be one of the powerful tools to study the Cosmic Dawn \citep[e.g.][]{Furlanetto2006CosmologyUniverse2,2023PASJ...75S...1S, 10.1093/pasj/psac015}. 
The spatial average and fluctuations of the 21cm signal evolved as the result of radiation from the first luminous objects. For example, the absorption of the 21cm global signal indicates coupling between the HI spin temperature and gas temperature via Lyman-$\alpha$ emission from the first stars (known as ``Wouthyusen-Field (WF) effect'' \citep{1952AJ.....57R..31W,1959ApJ...129..536F}).

There are many 21cm global signal instruments; EDGES \citep{2012RaSc...47.0K06R,2018Natur.555...67B}, BIGHORNS \citep{2015PASA...32....4S}, SARAS2 \citep{2022MNRAS.513.4507B}. In particular, the EDGES have continuously investigated the 21cm global signal, and various methods are developed \citep[e.g.][]{2017ApJ...835...49M,2022MNRAS.517.2264M}. The high-band data are used to constrain the galaxy evolution at the epoch of reionization \citep{2017ApJ...847...64M,2018ApJ...863...11M,2019ApJ...875...67M}. In 2018, the EDGES low band has reported a strong absorption line at $z=17.8$, which cannot be explained without new physics or unknown systematic error 
\citep{2018Natur.555...67B}.
The absorption line at high redshift indicates efficient structure formation at high redshift. This might indicate the high star formation rate and emission from the first stars \citep{Madau2018}. For example, the role of Pop III stars is investigated in \cite[e.g.][]{2020MNRAS.496.1445C} using the observed EDGES signal. The reported EDGES-low signal is also used for constraining the dark matter model \citep[e.g.][]{2020MNRAS.497.3393R}. 
One can also constrain primordial fluctuations, particularly on small scales \citep{Yoshiura2018, Yoshiura2020PhRvD.101h3520Y,Minoda2022PhRvD.105h3523M}, since the amplitude of density fluctuations on small scales is relevant to the structure formation process through which the 21cm signal can be affected\footnote{

Other probes of primordial power spectrum on small scales have been discussed such as primordial black holes \citep{2009PhRvD..79j3511B, 2009PhRvD..79j3520J, 2019PhRvD.100f3521S}; ultra-compact minihalos \citep{2012PhRvD..85l5027B,2018PhRvD..97b3539N, 2018JCAP...01..007E}; CMB spectral distortion \citep{1994ApJ...430L...5H,2012MNRAS.425.1129C,2012ApJ...758...76C,2013JCAP...06..026K,2014JCAP...10..046C,2016PhRvD..94b3523C,2017JCAP...11..002K}; supernovae lensing \cite{2016MNRAS.455..552B}; luminosity function of high-$z$ galaxies  \citep{Yoshiura2020PhRvD.101h3520Y}; dark matter substructure \citep{2022PhRvD.106j3014A}; 21cm fluctuations \citep{2013JCAP...10..065K,2018JCAP...02..053S,2017JCAP...05..032M}; 21-cm signal from minihalos \cite{2018JCAP...02..053S}; 21cm forest \citep{2014PhRvD..90h3003S} and Lyman-$\alpha$ forest \citep{2011MNRAS.413.1717B,2015JCAP...11..011P}, reionization history \citep{2023arXiv230409474M} and so on.
}. 

A follow-up observation with the SARAS3 has suggested the non-existence of such a strong absorption signal reported by EDGES \citep{2022NatAs...6..607S} and accordingly the high-$z$ astrophysics has been constrained \citep{2022NatAs...6.1473B}.
However, they only examined the absorption signal with the same shape as EDGES, and did not rule out the existence of absorption lines at the same redshift.

In addition to the global signal, one can also use the 21cm power spectrum to test the EDGES discovery. 
The observation of the 21cm power spectrum at the Cosmic Dawn has been actively operated for the last 10 years such as MWA \citep{Ewall-Wice2016, Yoshiura2021MNRAS.505.4775Y}, LOFAR \citep{Gehlot2019}, AARTFAAC \citep{10.1093/mnras/staa3093}, OVRO-LWA \citep{Eastwood2019, 2021MNRAS.506.5802G}, and  LWA-SV \citep{2020JAI.....950008D, 2021JAI....1050015D}. However, the current power spectrum results at the Cosmic Dawn have not constrained the models which can explain the strong absorption. 

Previous studies raised several possibilities of systematics in the EDGES data \citep{Hills2018ConcernsData2,2020MNRAS.492...22S,Singh2019TheSpectrum2,2019MNRAS.489.4007S}. Indeed any global signal instruments suffer from a number of systematics such as, for example, the ionosphere coupling with nonuniform foregrounds \citep{2021MNRAS.503..344S,2022MNRAS.515.4565S}. Detailed modeling and validation of instrumental beam response are mandatory \citep{2021AJ....162...38M, 2023MNRAS.521.3273S} even at the horizon \citep{2021ApJ...923...33B},  otherwise the biased signal will be recovered \citep[e.g.][]{2022MNRAS.515.1580S}. Apparently, simultaneous modeling of the foreground with antenna responses can mitigate the structured residuals \citep{2021MNRAS.506.2041A,2022arXiv221110448P}. Thus a prior antenna design should be required \citep{2022MNRAS.509.4679A}. It is also crucial to identify the unknown systematic error using Bayesian evidence to avoid biased results \citep{2022PASA...39...52S}. In the future, 21cm global signal observation with multiple instruments  \citep[REACH,][]{2022NatAs...6..984D, 2022arXiv221207415S} and observation from the far side of the moon \citep{2021arXiv210305085B,2021arXiv210308623B} can correctly remove/mitigate the systematics. Multiple approaches to measure the 21cm global signal have been now proposed \cite[e.g.][]{2020MNRAS.499...52M,2022PASA...39...18T,2022PASA...39...60P,2023ApJ...945..109Z}. Thus, in the upcoming years, a large amount of data observed with different instruments will be available. 

We here mention that the Cosmic Dawn is now receiving a great deal of attention from astronomers. Recent observations with the James Webb Space Telescope (JWST) found the most distant galaxies \citep[e.g.][]{2023MNRAS.518.4755A,2023MNRAS.519.1201A,2023MNRAS.518.6011D,2022ApJ...940L..55F,2023ApJ...942L...9Y,2022ApJ...940L..14N,2022ApJ...938L..15C,2022arXiv220802794N,Harikane2023,haro2023spectroscopic,2023ApJ...943L...9Z}. They revealed unexpected star formation rates during the Cosmic Dawn. The result suggests a high star formation rate and a top-heavy initial mass function. It would be important to test such scenarios with a different probe, and indeed the 21cm line can be a key alternative observable to reveal astrophysical processes during the Cosmic Dawn. 

It is therefore imperative to develop software to study the 21cm global signal toward a Bayesian analysis with future observations. For example, the 21CMMC \citep{2015MNRAS.449.4246G,2017MNRAS.472.2651G,2018MNRAS.477.3217G} has been developed for the Bayesian MCMC analysis while the change of cosmological parameters makes the performance slow. 
In \cite{2021MNRAS.507.2405C}, they have developed the MCMC package (CosmoReionMC) and constrained the astrophysical and cosmological parameters including the spectral index and the amplitude of the primordial power spectrum using the combination of data from the CMB observed by Planck, quasar absorption spectra and 21cm global signal. 
Very recently, \cite{2022arXiv221214064N} performed the MCMC analysis with parameters describing the bump in the primordial power spectrum and 2 astrophysical ones. However, the prediction of a 21cm signal needs high computational costs to calculate the complicated process with many astrophysical and cosmological parameters in three dimensions. For a quick and accurate prediction of a 21cm signal, emulators have been developed (\cite{2017ApJ...848...23K,GLOBALEMUBevins,2020MNRAS.495.4845C,2022ApJ...930...79B, 2018MNRAS.475.1213S}) although the running and running of running parameters of primordial power spectrum, which is a common way of describing the detailed scale-dependence, especially on small scales, are not taken into account. 

In this work, we build an artificial neural network (ANN) based emulator so that we can efficiently predict the 21cm global signal with multiple astrophysical parameters and the ones to characterize the primordial fluctuations such as the spectral index and the running parameters. We demonstrate parameter constraints with Bayesian analysis using the developed emulator and publicly available EDGES low-band data.

This paper is organized as follows. In the next section, we describe the model of the 21cm signal used to create the training and test data and the algorithm of emulation. Section 3 contains the description of our Bayesian analysis. In Section 4, we give the results of the Bayesian analysis and discuss how the assumptions of the foreground, noises, and frequency range affect parameter constraints. Comparisons of models suggested by EDGES and JWST are also discussed. We summarize this paper in the final section. Through this paper, we fix the cosmological parameters as $\Omega_{\rm m}=0.316$, $\Omega_{\Lambda}=1-\Omega_{\rm m}$, $\Omega_{\rm b}=0.0491$, $\rm H_0=67.27$, $\sigma_8=0.831$ \citep{2016A&A...594A..13P}.

\section{Methods: Modelling 21cm Signal, Emulator and Luminosity Function}

This section describes the modeling of the 21cm line brightness temperature, the high-$z$ ultra-violet luminosity function (UVLF), and the ANN-based emulator.

\subsection{21cm brightness temperature}

The 21cm line is observed as emission/absorption against the radio background such as the CMB, and then the 21cm line brightness temperature is given as \citep[e.g.][]{Furlanetto:2006jb}
\begin{eqnarray}
\delta T_{\rm b} \approx 27 x_{\rm HI} (1+\delta) \left( \frac{T_{\rm S}-T_{\rm \gamma}}{T_{\rm S}} \right) \left( \frac{1+z}{10}\right)^{\frac{1}{2}} \rm [mK],
\end{eqnarray}
where $T_{\rm S}$ is the spin temperature, $x_{\rm HI}$ is the neutral fraction of hydrogen gases, $\delta$ is matter over-density and $T_{\gamma}=2.725/(1+z)$ is the temperature of CMB. Note that we do not consider the enhanced radio backgrounds \citep[e.g.][]{2018ApJ...858L..17F,2019MNRAS.486.1763F,2020MNRAS.499.5993R,2020MNRAS.492.6086E} proposed to reproduce the very strong absorption signal reported in \cite{2018Natur.555...67B}.

We used a modified version of 21cmFASTv2\footnote{There are several recent updates on the 21cmFAST such as the inhomogeneous Lyman-Werner and relative-velocity feedback \citep{2022MNRAS.511.3657M}.}  \citep{2011MNRAS.411..955M, Park2019}, which is similar to the methodology 
used in \cite{Yoshiura2018,2023arXiv230409474M}, for evaluating the 21cm line brightness temperature. Specifically, we added the running $\alpha_{\rm s}$ and the running of running $\beta_{\rm s}$ to the primordial power spectrum. The detailed method used in the 21cmFAST for solving the ionization and evolution of spin temperature is well described in \cite{2011MNRAS.411..955M, Park2019}. We here describe the method briefly. 
The initial matter power spectrum will be used for creating an initial high-resolution matter density map and estimation of the halo mass function\footnote{
The choice of halo mass function, the stellar population synthesis, and cosmology can affect the 21cm global signal with a few - tens of mK \citep{2021MNRAS.504.1555M}.
}. The three-dimensional matter density map of each redshift bin is evaluated using 2LPT from the initial matter distribution with high resolution. The ionization is solved by comparing the number of neutral baryons and recombination with ionizing photons \citep{2014MNRAS.440.1662S}. The heating of IGM is caused by the X-ray emission\footnote{
Note that additional heating processes such as the Lyman-$\alpha$ heating produced by the Lyman-$\alpha$ photons can change the shape of global signal \citep{2020PhRvD.101h3502V,2020MNRAS.492..634G,2021MNRAS.503.4264M}.
}, calculated as angle-averaged specific intensity, with the optical depth being taken into account. The Lyman-$\alpha$ background is originated from the X-ray excitation of the neutral hydrogen atom and direct stellar emission of the Lyman series. The 21cm signal is evaluated using the maps of neutral fraction, spin temperature, and matter density. The global signal (i.e. sky averaged) 21cm signal is calculated using the resultant 21cm brightness temperature maps. 

The emission from luminous objects is controlled by several astrophysical parameters, which we list following \cite{Park2019}; the fraction of gas converted into stars in a halo of mass $10^{10} M_{\odot}$, $f_{\rm star}$, the power-law index of $f_{\rm star}$ as a function of halo mass, 
$\alpha_{\rm star}$, a fraction of time scale of the star formation against Hubble time, $t_{\rm star}$, the escape fraction of ionizing photons of a halo of mass $10^{10} M_{\odot}$, $f_{\rm esc}$, the power-law index of $f_{\rm esc}$ as a function of halo mass, $\alpha_{\rm esc}$, turn over mass, $M_{\rm turn}$, which is the lowest halo mass to host the efficient star formation,
X-ray luminosity per star formation, $L_{\rm X}$, and minimum X-ray energy, $E_0$.

In addition to astrophysical parameters, the small-scale primordial power spectrum can also play an important role in the evolution of the 21cm global signal. The primordial power spectrum can be parametrized as 
\begin{eqnarray}
\label{eq:P_prim}
P_{\rm prim} \propto \left(\frac{k}{k_{\rm ref}}\right)^{n_{\rm s} -1 + \frac{1}{2}\alpha_{\rm s} \ln({k/k_{\rm ref}}) + \frac{1}{6}\beta_{\rm s} \ln^2({k/k_{\rm ref}})  },
\end{eqnarray}
where $n_s$ is the spectral index, $\alpha_s$ and $\beta_s$ are the running and the running of the running parameters. $k_{\rm ref}=0.05 {\rm Mpc^{-1}}$ is the reference scale. In \cite{Yoshiura2018, Yoshiura2020PhRvD.101h3520Y}, the effects of $\alpha_{\rm s}$ and $\beta_{\rm s}$ on the 21cm global signal are investigated. The enhancement of the matter power spectrum owing to the positively large value of $\alpha_{\rm s}$ and $\beta_{\rm s}$ makes the structure formation faster. As a result, the redshift of the 21cm absorption signal is shifted to earlier. Thus the primordial power spectrum is also important in predicting the 21cm global signal and we vary $n_s$, $\alpha_s$, and $\beta_s$ as well as the astrophysical parameters in the following analysis.

We run the 21cmFAST for roughly 49000 different parameter sets in the PC cluster of CfCA and store the global 21cm line signal. The simulation was performed in a cosmological volume of $128{\rm Mpc^3}$, $192^3$ cells for initial condition and $64^3$ cells for ionized and temperature maps\footnote{
We tested various sets of resolutions and find that the resolution does not significantly affect the evolution of the 21cm global signal.
}. 
We divide the redshift into 53 bins over the range of $6<z<30$. 
Parameters are randomly selected from a range motivated by \cite{Park2019}. The ranges of parameters of the primordial power spectrum ($n_{\rm s}$, $\alpha_{\rm s}$, and $\beta_{\rm s}$) are consistent with the 5 $\sigma$ confidence level of the constraints from \cite{2020A&A...641A..10P}. The ranges of parameters are listed in Table~\ref{tab:1}. The set of parameters and the resultant 21cm global signal is used for training our emulator as described below. We will use roughly 44000 data as training data and the rest as test data. 

\begin{table}
\caption{List of parameters and the range used to create training data sets.}
\begin{tabular}{lll}
   & lower limit & higher limit \\
   \hline
$n_{\rm s}$       & 0.9385          & 0.9865 \\
$\alpha_{\rm s}$       & -0.048          & 0.052 \\
$\beta_{\rm s}$      & -0.055          & 0.075 \\
$\log_{10} f_{\rm star}$       & -3.0          & 0.0 \\
$a_{\rm star}$       & -0.5          & 1.0 \\
$\log_{10} f_{\rm esc}$      & -3.0          & 0 \\
$a_{\rm esc}$      & -1.0          & 0.5 \\
$\log_{10} M_{\rm turn}$      & 8          & 10 \\
$t_{\rm star}$      & 0          & 1 \\
$\log_{10} L_{\rm X}$      & 38          & 42 \\
$E_{\rm 0}$      & 100          & 1500 \\
\hline
\end{tabular}
\label{tab:1}
\end{table}

\subsection{Emulator}

\begin{figure}
\centering
\includegraphics[width=8cm]{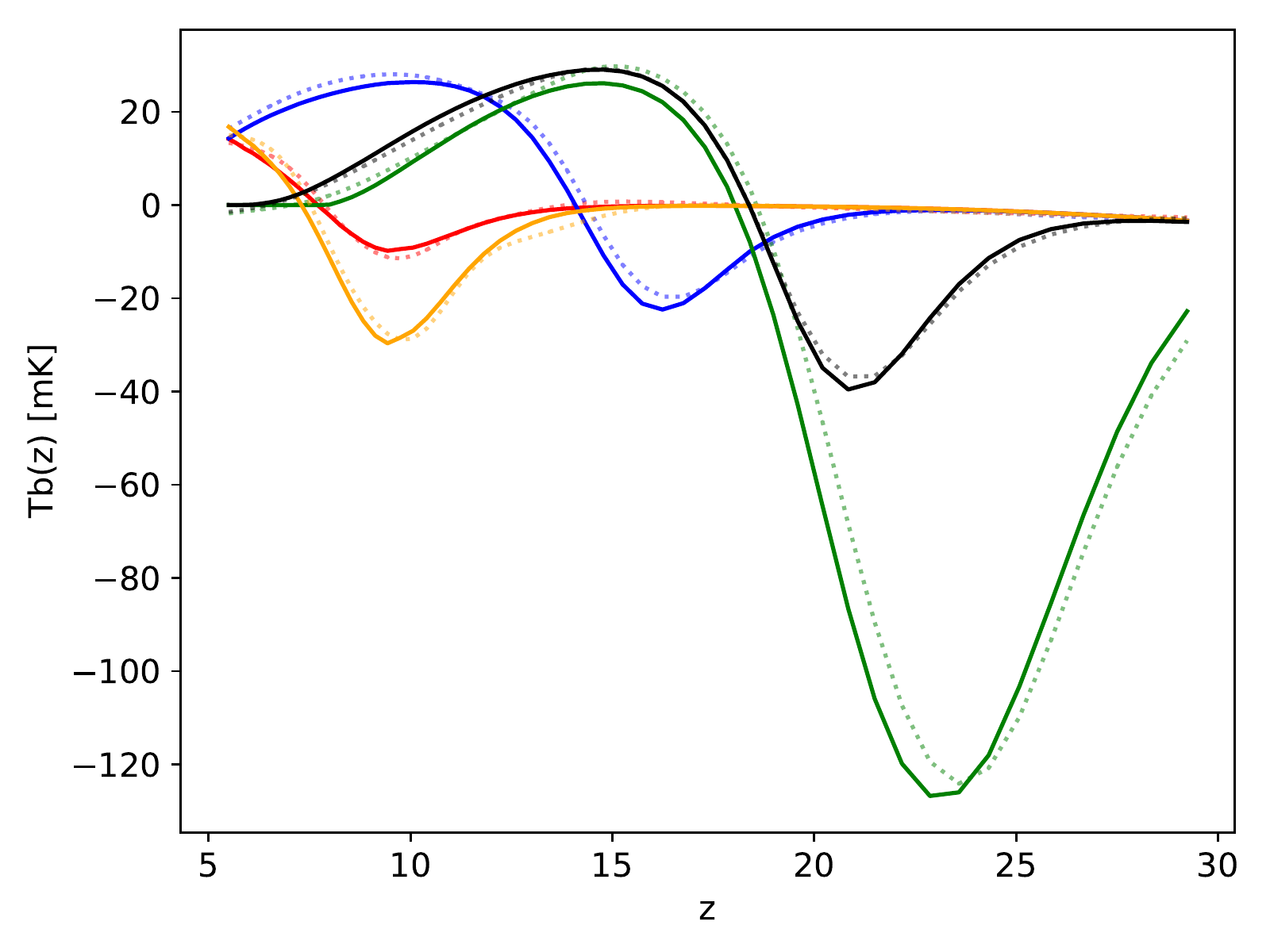}
\caption{Examples of 21cm global signals. Solid lines are randomly taken from the test data set. The dotted lines are predicted using the emulator we developed. }
\label{fig:1}
\end{figure}

As the emulator of a 21cm global signal, we employ an ANN-based architecture which is similar to some previous works \citep[][]{2020MNRAS.495.4845C,GLOBALEMUBevins,2022ApJ...930...79B}. We perform the optimization of ANN models using 12 different architectures\footnote{
The 12 models are the combinations of the number of hidden layers (2, 3, and 4) and the number of neurons at the hidden layers (4,8,16, and 32).
}.
The hidden layers have a Tanh activation function except for the last layer which has a linear activation. The input layer of the network for the global signal has 11 inputs (8 astrophysical parameters, 3 primordial power spectrum parameters), and 1 redshift index. The redshift indices (from 1 to 53) correspond to the output redshift bins of 21cmFAST. The network yields one output corresponding to the brightness temperature at the redshift index. The loss function used to optimize the network is Mean Squared Error. We use the batch size of 53, ADAM optimizer, and learning rate of 0.002. We train the network for 100 epochs using the all training data set. As quantitative evaluators of the accuracy of our network, we employ root mean squared error (RMSE) as in \cite{GLOBALEMUBevins,2020MNRAS.495.4845C}. The RMSE for a test data set is given as
\begin{eqnarray}
RMSE = \sqrt{\frac{1}{N}\sum_{j=0}^{N} \left(
\delta T_{\rm sim}(z_j) - \delta T_{\rm pred}(z_j)
\right)^2},
\end{eqnarray}
where $N$ is number of redshift bins, $z_j$ is redshift at $j$-th bin, $\delta T_{\rm sim}$ is simulated 21cm brightness temperature from 21cmFAST and $\delta T_{\rm pred}$ is 21cm signal predicted using the emulator. 
As a result, the architecture consists of 1 input layer, 4 hidden layers with 32 neurons for each layer, and 1 output layer, achieving the mean RMSE of 2.45 mK which is the smallest among the other architectures. The figure~\ref{fig:1} shows examples of the 21cm global signal predicted by our emulator.
The accuracy might be improved further by optimization of training parameters and the architecture. However, this RMSE is more or less consistent with the previous work and lower than the expected thermal noise error of the EDGES data at a frequency bin. Thus, we consider that the current emulator is accurate enough for this work. 

Note that the redshift bins of our training data sets do not correspond to the frequency bins of the EDGES data one by one. Thus, in the Bayesian analysis, we predict the 21cm brightness temperature by linearly interpolating the values to evaluate the 21cm signal at the frequency bins of the EDGES data.

\subsection{Luminosity function}

Using the parameters used in the 21cmFAST, we can evaluate the high-$z$ UVLF following \cite{Park2019}. The star formation rate (SFR) is described as
\begin{eqnarray}
\dot{M}_* (M_{\rm h}, z) = M_{\rm *} \frac{M_{\rm h}}{t_{\rm star}H_z^{-1}},
\end{eqnarray}
where $H^{-1}_z$ is Hubble time at the redshift $z$ and the stellar mass $M_{\rm *}$ is given as
\begin{eqnarray}
M_* (M_{\rm h}, z) = f_{\rm star, m} \left(\frac{\Omega_{\rm b}}{\Omega_{\rm m}}\right) M_{\rm h},
\end{eqnarray}
where $M_{\rm h}$ is halo mass and the $f_{\rm star, m}$ is less than 1 and given as $f_{\rm star, m}$ = $f_{\rm star} M^{\alpha_{\rm star}}_{\rm h,10} $. $M_{\rm h,10}$ is halo mass normalized by $10^{10} M_{\odot}$. The SFR is converted to rest-frame UV luminosity by the conversion factor $K_{\rm UV}$ = $1.15 \times 10^{-28} M_{\odot}~ {\rm yr^{-1} /ergs~ s^{-1} Hz^{-1}}$ \citep[assuming the Salpeter IMF,][]{2016MNRAS.460..417S}. The UV magnitude $M_{\rm UV}$ is derived from the luminosity by following AB magnitude relation \citep{1983ApJ...266..713O}. Finally, the 
UVLF is given as 
\begin{eqnarray}
\phi(M_{\rm UV}) = f_{\rm duty}\frac{d n }{d M_{\rm h}} \left|\frac{dM_{\rm h}}{d M_{\rm UV}}\right|,
\end{eqnarray}
where $f_{\rm duty} = \exp(-M_{\rm turn}/M_{\rm h})$ is a suppression factor for star formation in small halos. The halo mass function ${d n }/{d M_{\rm h}}$ is calculated following \citep{2013A&C.....3...23M} 
with the parametrization for the primordial power spectrum\footnote{We here assume the amplitude of the primordial power spectrum at the reference scale is $2.207 \times 10^{-9}$ \citep{2016A&A...594A..13P}. } \eqref{eq:P_prim} as in e.g. \cite{2020PhRvD.102h3515Y},  using the top-hat window function.  

Previous works \citep{Park2019,2020PhRvD.102h3515Y} have shown that the UVLF at $z \le 10$ is a powerful quantity to constrain the astrophysical parameters and primordial power spectrum. However, when we perform the Bayesian analysis using the EDGES data corresponding to $z>13$, we do not include the UVLF in the likelihood. We calculate the UVLF at $z=12$ and $z=16$ only to discuss the implication of EDGES to the recent JWST results.

\section{Method: Data Analysis}
\label{sec:3_method}

Using our emulator to evaluate the 21cm global signal, we perform the Nested sampling Bayesian analysis to constrain the astrophysical parameters with several assumptions for foreground model, systematics, and appropriate noise level. To this end, we use the publicly available EDGES low-band data \citep{2018Natur.555...67B}\footnote{
\url{https://loco.lab.asu.edu/edges/edges-data-release/}
}. 
The data contains an integrated sky spectrum from 51 MHz to 99 MHz. For the model of foreground, systematics, and noise, we adopt those motivated in \cite{2020MNRAS.492...22S}. We refer the reader to the reference for a more detailed discussion of the models. We describe the methods below following \cite{2020MNRAS.492...22S}.

We define a Gaussian log-likelihood function\footnote{In \cite{2022arXiv220404491S}, they argued that the generalized normal likelihood function would be an appropriate likelihood function if the noise model is unknown. In this work, we assume the noise follows Gaussian distribution. Therefore a Gaussian log-likelihood would be suitable in our analysis.} for the model $M$ as, 
\begin{eqnarray}
    \ln \mathcal{L}({\bf \theta},M) = -\frac{n}{2} \ln(2\pi) - \frac{1}{2}\sum^n_i \ln \sigma_i^2 -  \sum^n_i \frac{1}{2}  \frac{(D_i - {m}_i ({\bf \theta}))^2 }{\sigma^2_i} ,
\end{eqnarray}
where $n$ is the number of frequency channels, $\sigma_i$ is the  noise level of EDGES data at  $i$-th frequency channel, $D_i$ is the EDGES low-band data at channel $i$, ${m}_i$ is the component of our model $M$ at channel $i$ and ${\bf \theta}$ is a vector of our model parameters. 
Based on Bayes's theorem, the posterior distribution function of the parameters is given as
\begin{eqnarray}
   P({\rm \theta}|{\mathbf{D}}, {M})= \frac{\mathcal{L} ({\mathbf{\theta} },{M})\pi ({\bf \theta},{M})}{\mathcal{Z}} \,,
\end{eqnarray}
where $\pi$ is the parameter's prior probability distribution and the Bayesian evidence, $\mathcal{Z}$, is given as
\begin{eqnarray}
   \mathcal{Z} = \int \mathcal{L}({\bf \theta}) \pi ({\bf \theta})d^{x}{\bf \theta},
\end{eqnarray}
where $x$ is the number of dimensions of our parameter space. Comparing the value of $\mathcal{Z}$, we can perform model selection. If some two models $M_0$ and $M_1$ are equally probable a priori, $\ln\mathcal{Z}_0-\ln\mathcal{Z}_1>3$ indicates the model $M_0$ is more likely than $M_1$ \citep{Kass1995}. In this work, we use Polychord \citep{2015MNRAS.450L..61H,2015MNRAS.453.4384H} to calculate the Bayes evidence and obtain the posterior probability distribution of parameters. The accuracy of the evidence and posterior distribution can be improved by increasing the number of live points ${\rm n_{\rm live}}$ which is a parameter used in Polychord. We use ${\rm n_{\rm live}}=2000$ for evaluating the $\ln \mathcal{Z}$. For some models, we checked that the evidence values evaluated with ${\rm n_{\rm live}}=4000$\footnote{To improve the accuracy of the posterior distribution, we use the results with ${\rm n_{\rm live}}=4000$ in figure 4-12.} are consistent with the values calculated with ${\rm n_{\rm live}}=2000$.

The measured sky signal is dominated by the Galactic synchrotron radiation and can be described with a smooth function of the frequency. 
We, therefore, use the $N$-th log polynomial function as our foreground model
\begin{eqnarray}
    T_{\rm FG} = 10^{\sum_{i=0}^{N}d_i \log_{10} (\nu/\nu_0)^i} 
    \ [\mathrm{K}],
\end{eqnarray}
where $d_i$ are the model parameters used in our fitting and $\nu_0=75 \rm MHz$. Foregrounds are expected to be well smooth \citep{1999A&A...345..380S} and described with at least 3rd order polynomial function \citep{2010PhRvD..82b3006P}. The ionosphere absorption and miscalibration of the beam and instrumental gain might be sources of further fluctuation of the spectrum as mentioned above. 
A higher-order polynomial function could allow more freedom in modelling the modulated foreground but also could lead to overfitting, so the number of polynomial orders should be determined carefully.
We, therefore, vary the order of polynomials from 5th to 9th. 
It is worth mentioning that several advanced methods have been rapidly developing  to analyze the foreground-dominated 21cm observation \citep[e.g.][]{2021ApJ...908..189B, 2021MNRAS.502.4405B}. 
However, in this study, we use a traditional polynomial function to fit the foregrounds for simplicity.

As found in previous works \citep[e.g.][]{Hills2018ConcernsData2, Singh2019TheSpectrum2, 2020MNRAS.492...22S}, systematics with sinusoidal shape might exist in the EDGES low-band data. The sinusoidal systematics, for example, can arise due to inaccurate correction of instrumental beam response. Thus, we employ the function below to model the systematics,
\begin{eqnarray}
    T_{\rm cal}(\nu) = \left(\frac{\nu}{\nu_0}\right)^b \{10^{a_0} \sin{(2\pi\nu/P)}+10^{a_1} \cos{(2\pi\nu/P)}\}
    \ [\mathrm{K}],
    \label{eq:sin}
\end{eqnarray}
where $a_0$, $a_1$, $b$ and $P$ are free parameters of the systematics.

The thermal noise of the global signal observation can be given as $T_{\rm rms}(\nu)  = [{T_{\rm sky}(\nu)  + T_{\rm rec}}]/\sqrt{w (\nu) \Delta t \Delta \nu }$. Here the sky temperature $T_{\rm sky}$ is the integrated sky spectrum of EDGES. According to \cite{2018Natur.555...67B}, we can assume the receiver temperature $T_{\rm rec}$ to be
200~K, the channel width of $\Delta \nu=0.390625$ MHz and the effective integration time of $\Delta t=$107 hours. The data at higher than 87 MHz suffer from the strong contamination of RFI\footnote{
See the Extended Data Figure 7 in \cite{2018Natur.555...67B}.
}. Thus, the effective integration time is corrected using normalized channel weights, $w$ and the normalized weight has low values at $\nu>87$ MHz. For comparison, we perform the Bayesian analysis using the data only at $\nu<87$~MHz and all frequency data ($51 < \nu < 99$ MHz). Regrading the noise level, 
if there are additional noise sources such as calibration error and polarized foregrounds as discussed in \cite{2020MNRAS.492...22S}, they can be larger than the thermal noise. 
We, therefore, define the noise level as $\sigma_i = A_{\rm{n}} T_{\rm rms}(\nu_i) + T_{\rm wn}/\sqrt{w(\nu_i)}$,
and we perform the Bayesian analysis using the theoretical noise model ($A_{\rm{n}}=1$, $T_{\rm wn}=0$ K) and enhanced noise model ($A_{\rm{n}}=2.25$, $T_{\rm wn}=0.015$ K) for comparison. The enhanced noise model is motivated by the maximum posterior parameters of the highest evidence models in \cite{2020MNRAS.492...22S}. 
The parameters and the ranges assumed in the Bayesian analysis for the foreground, the systematics, and the noise level are summarized in Table~\ref{tab:2}.

In the Bayesian analysis, the measured signal should be compared to the sum of the 21cm global signal predicted using the emulator, foreground, the sinusoidal systematics, and the thermal noise.

\begin{table}
\caption{List of parameters and the range used for the Bayesian analysis. For the astrophysical and cosmological parameters of the emulator, the prior range is listed in Table~\ref{tab:1}.
}
\begin{tabular}{lll}
  param & lower limit & higher limit \\
   \hline
$d_0$       & -5          & 5 \\
$d_1$       & -3          & -2 \\
$d_2-d_9$       & -100          & 100 \\
$a_0$       & -10.0          & 2.0 \\
$a_1$       & -10.0          & 2.0 \\
$P$       & 10.0          & 15.0 \\
$b$       & -4.0          & 4.0 \\
\hline
\end{tabular}
\label{tab:2}
\end{table}

\section{Results and Discussion}

\begin{figure}
\centering
\includegraphics[width=8cm]{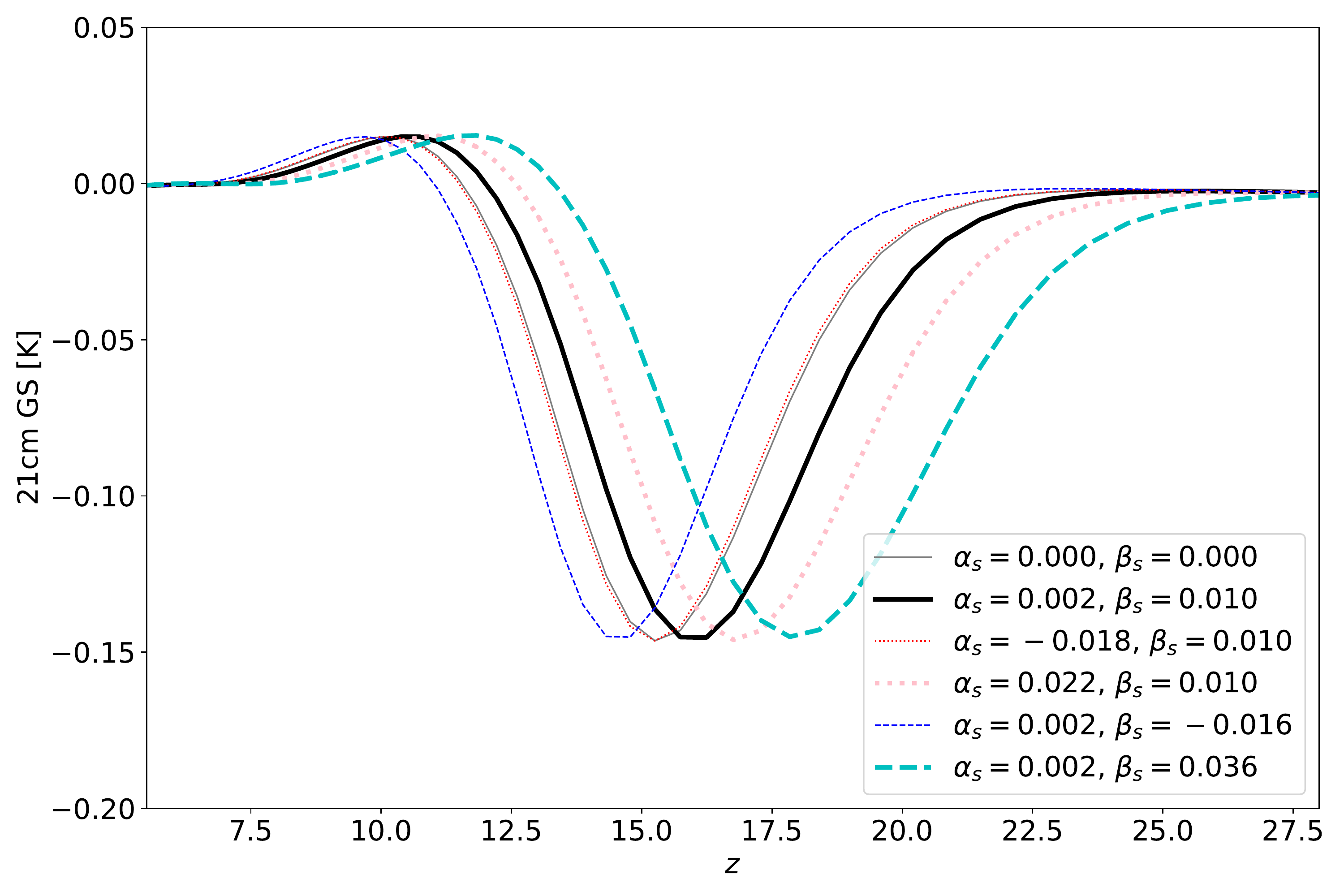}
\caption{Impact of running parameters on the 21cm global signal. The signals are calculated using the emulator. The thick solid line is our fiducial model ($\alpha_{\rm s}=0.002$ and $\beta_{\rm s}=0.01$). }
\label{fig:runningeff}
\end{figure}

We first check the effect of running parameters on the 21cm global signal in figure \ref{fig:runningeff}. The figure shows the 21cm global signal calculated using the emulator developed in this work. The thick solid line is the model with $\alpha_{\rm s}=0.002$ and $\beta_{\rm s}=0.01$ corresponding to the mean values in Planck 2018 \citep{2020A&A...641A..10P}. The figure clearly shows that the higher (lower) values of $\alpha_{\rm s}$ and $\beta_{\rm s}$ shift the 21cm 
absorption trough to the higher (lower) redshift. This shift is caused due to the enhancement (reduction) of the initial matter power spectrum which 
affects the halo mass function \citep[e.g.][]{Yoshiura2020PhRvD.101h3520Y, 2023arXiv230409474M}. 

\begin{figure*}
\centering
\resizebox{180mm}{!}{
\includegraphics[width=4cm]{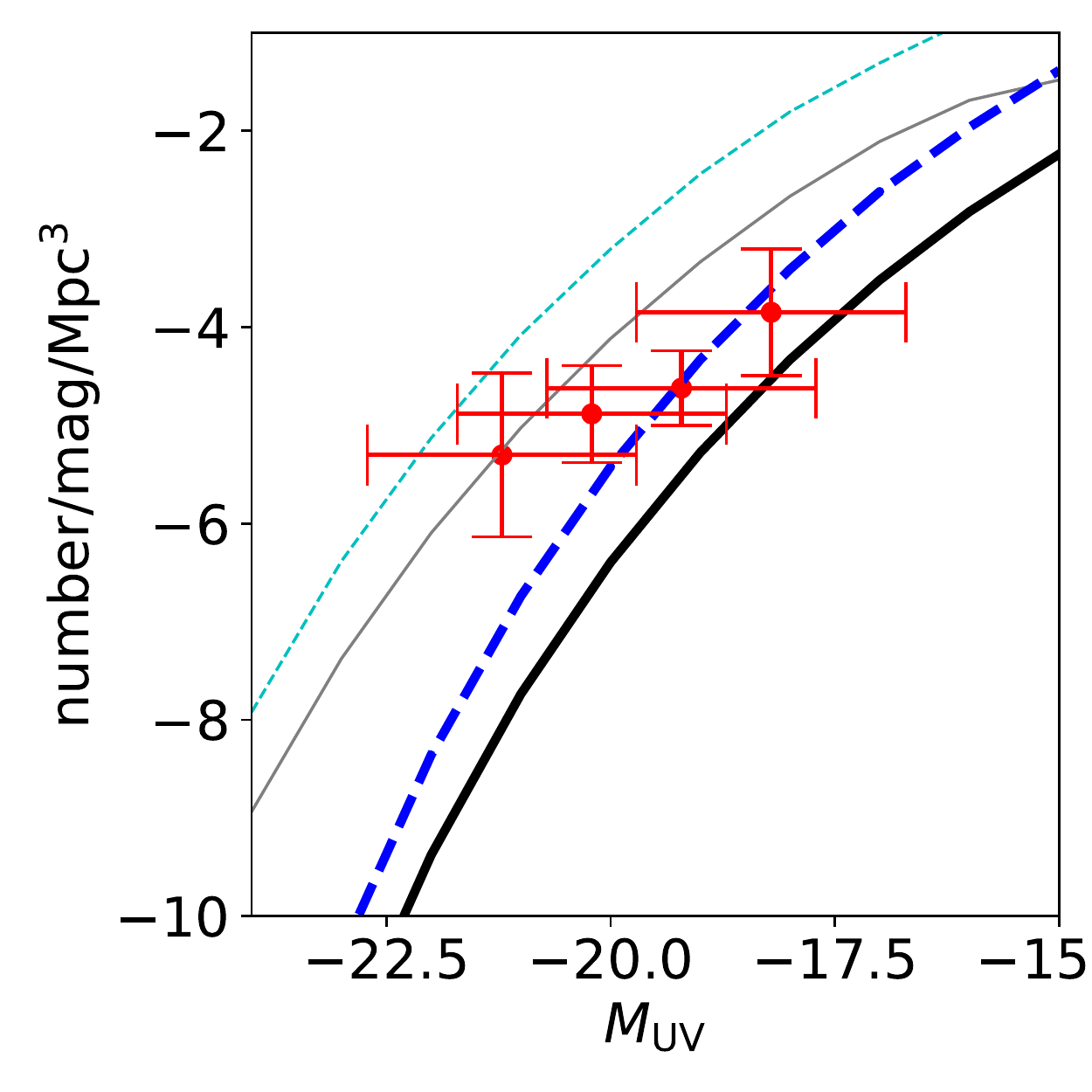}
\includegraphics[width=4cm]{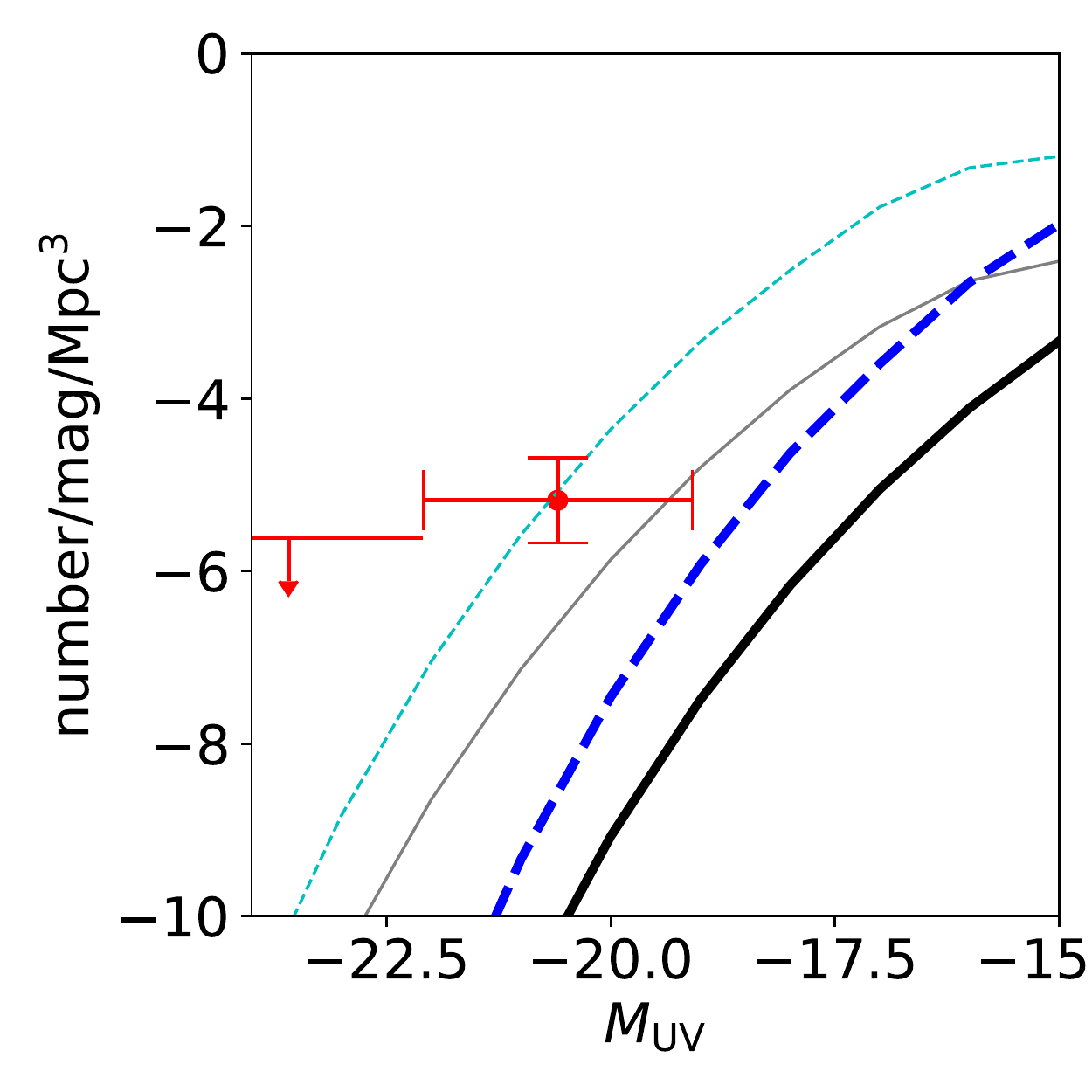}
\includegraphics[width=6cm]{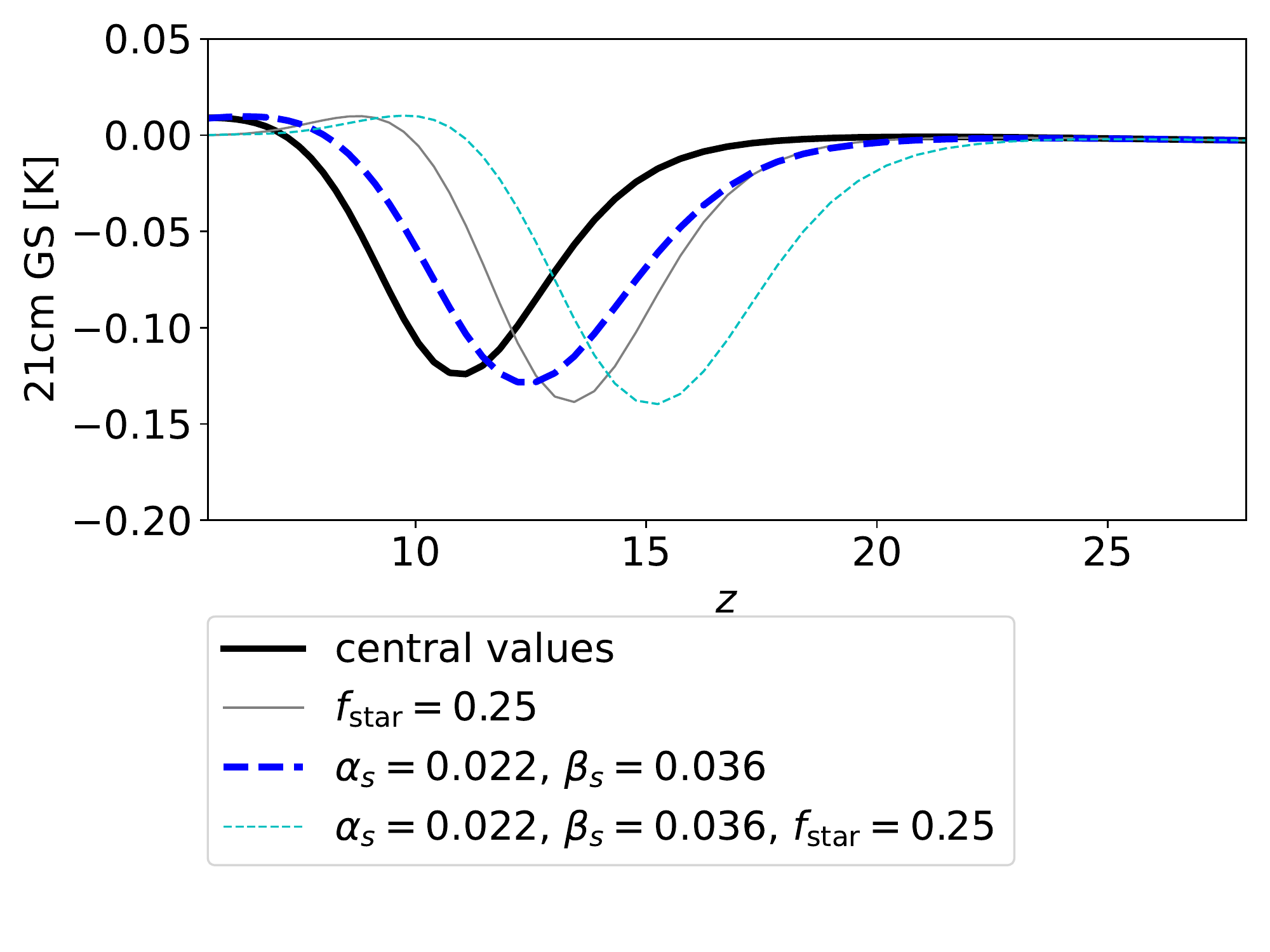}
}
\caption{Examples of 21cm global signal and UVLF. The thick solid line is the model with all parameters being central values of the prior range. The thin solid line has the same parameters as the thick solid line except $f_{\rm star}$ of 0.25. Dashed lines are the model with $\alpha_s=0.022$ and $\beta_s=0.036$. }
\label{fig:compGSLF}
\end{figure*}

We next see the connection between the 21cm global signal and UVLF. Figure \ref{fig:compGSLF} shows the 21cm global signal and UVLF at $z=$12 and 16. The thick solid line is
calculated using the central parameter values of the prior range (Table~\ref{tab:1}). 
With the central values for the parameters, we find the 21cm global signal has an absorption trough at $z=11$. The corresponding UVLF of $z=12$ and $z=16$ are not consistent with observational results of \cite{Harikane2023}. 

Now we demonstrate how the 21cm global signal and UVLF are affected by varying these cosmological and astrophysical parameters. Among the parameters listed in Table~\ref{tab:1}, the effect of $f_{\rm star}$ on the high-$z$ UVLF, which is measured in JWST, is particularly important. Its central value in the prior range is $f_{\rm star}=0.03$, and we show the case of $f_{\rm star}=0.25$ with the thin solid line in figure \ref{fig:compGSLF}. In this case, the UVLFs at $z=12$ and $16$ overlap with the JWST results within 1$\sigma$. Further enhancement might be required to explain the JWST best-fit value of UVLF at $z=16$, but such a parameter can easily conflict with the observation of galaxies at lower redshifts. 

Instead of increasing $f_{\rm star}$, larger running parameters $\alpha_{\rm s}$ and $\beta_{\rm s}$ can increase the UVLF. We show the results with $\alpha_{\rm s}=0.022$, $\beta_{\rm s}=0.036$, and the other parameters remaining the central values in the prior range as the thick dashed line in figure \ref{fig:compGSLF}. This result is fitted well to the UVLF data points at $z=12$, but the discrepancy between the UVLF at $z=16$ still remains. Additionally, increasing both of the running parameters and $f_{\rm star}$ is also shown with the thin dashed line. The UVLF at $z=16$ can well explain the observation of JWST, but the UVLF at $z=12$ fails. The discrepancy may require modified models such as the redshift evolution of astrophysical parameters, emission from Pop III stars, and UV luminosity calculated using the top-heavy IMF and so on \citep{Harikane2023}. Since our main focus is the 21 cm global signal, we do not include the UVLF in the likelihood of the Bayesian analysis which we will see in the following subsections.

\subsection{Comparing All Cases}

Our emulator is constructed with 11 parameters. We demonstrate the emulator with the EDGES low-band data which covers the frequency range of the 21cm line before the cosmic reionization. Thus, we fix the escape fraction as the $\log_{10} f_{\rm esc}=-1.5$ and $\alpha_{\rm esc}=-0.25$ since these parameters are mainly relevant to the reionization. We also consider a model with fixing the parameters for the primordial power spectrum as $n_{\rm s}=0.9665$, $\alpha_{\rm s}=0.0$, and $\beta_{\rm s}=0.0$. It would be worth mentioning that the model with $n_{\rm s}=0.9625$, $\alpha_{\rm s}=0.002$, and $\beta_{\rm s}=0.01$, which corresponds to the mean values of Planck constraint \citep{2020A&A...641A..10P}, has the absorption peak at higher redshifts than that of the model with $\alpha_{\rm s}=\beta_{\rm s}=0$ as shown as the thin solid line in figure~\ref{fig:runningeff}. 
However, the influence is minor compared to the effects of astrophysical parameters. We also examine the model without a 21cm signal. 

We assess the foreground models with orders of polynomials from 5th to 9th. The prior range of the parameters used in the fitting is listed in Table~\ref{tab:1} and \ref{tab:2}. We perform the analysis for multiple combinations of these models as listed in Table~\ref{tab:3}.

Regarding the systematics, we found that the model without the sinusoidal systematics, described as Eq.~\eqref{eq:sin}, has significantly larger RMS and lower $\ln \mathcal{Z}$. This might suggest the existence of sinusoidal systematics. Therefore we decided to use sinusoidal systematics in all models.

We perform the Bayesian analysis using all frequency bands (51-99MHz) or only lower bands (51-87 MHz) and use 2 different noise models. For clear discussion, we define 4 cases for the combination of the frequency band and noise model, namely FullBW(full bandwidth), HcutBW (high-frequency cut bandwidth), TN (theoretical noise, $A_{\rm n} = 1$, $T_{\rm wn} = 0$ K), and EN (enhanced noise, $A_{\rm n} = 2.25$, $T_{\rm wn} = 0.015$ K). In addition, depending on the model framework as to how we fix the parameters, we separate the models into 4 labelled as M1, M2, M3 and M4 (see Table~\ref{tab:3}). We also define 11 IDs for different combinations of 21cm signal and foregrounds (see Table~\ref{tab:3}). We refer, for example, to the model of M1-FG8 and FullBW-TN as FullBW-TN-M1-FG8.

\begin{table*}
\caption{List of model combinations used in the Nested sampling analysis. The Bayesian evidence values, $\ln \mathcal{Z}$, are also listed.
The largest Bayesian evidence in each systematic (frequency range and noise) model is highlighted in bold. We use the $\rm n_{\rm live}=2000$ to evaluate $\ln \mathcal{Z}$ listed in this table. }
\begin{tabular}{lllllll}
   &   &  & \multicolumn{2}{c}{$\ln \mathcal{Z}$ ($A_{\rm n} = 1$, $T_{\rm wn} = 0$) }&  \multicolumn{2}{c}{ $\ln \mathcal{Z}$ ($A_{\rm n} = 2.25$, $T_{\rm wn} = 0.015$) }\\
   \hline
   &  &  & FullBW-TN   & HCutBW-TN &  FullBW-EN  & HCutBW-EN  \\ 
   ID & 21cm params & foreground & (51$<\nu<$99 MHz)   & (51$<\nu<$87 MHz) &  (51$<\nu<$99 MHz)  & (51$<\nu<$87 MHz)  \\
   \hline
M1-FG5      & Fix $f_{\rm esc}$, $a_{\rm esc}$, $n_{\rm s}$, $\alpha_{\rm s}$, $\beta_{\rm s}$  & 5th order      &   -397.213   &   -32.5    & 199.679 & 157.974  \\
M1-FG6       & Fix $f_{\rm esc}$, $a_{\rm esc}$, $n_{\rm s}$, $\alpha_{\rm s}$, $\beta_{\rm s}$  & 6th order      &   -272.954   &    -24.5   & 216.410  & \textbf{158.030} \\
M1-FG7       & Fix $f_{\rm esc}$, $a_{\rm esc}$, $n_{\rm s}$, $\alpha_{\rm s}$, $\beta_{\rm s}$  & 7th order      &   -274.281   &     -22.5  & 215.952  & 157.560  \\
M1-FG8       & Fix $f_{\rm esc}$, $a_{\rm esc}$, $n_{\rm s}$, $\alpha_{\rm s}$, $\beta_{\rm s}$  & 8th order      &   -271.662   &    -22.2   &  216.101  & 157.692 \\
M1-FG9       & Fix $f_{\rm esc}$, $a_{\rm esc}$, $n_{\rm s}$, $\alpha_{\rm s}$, $\beta_{\rm s}$  & 9th order      &   -273.323   &  \textbf{-22.0}   & \textbf{216.770}  &  157.780\\
M2-FG5       & Fix $f_{\rm esc}$, $a_{\rm esc}$ & 5th order      &   -372.202   &    -31.4   & 200.155 & 157.682 \\
M2-FG9       & Fix $f_{\rm esc}$, $a_{\rm esc}$ & 9th order      &   -271.537   &     -22.9  & 216.038 & 157.817 \\
M3-FG5       & use all & 5th order      &   -327.27   &    -32.3   & 203.109 & 157.550 \\
M3-FG9       & use all & 9th order      &    \textbf{-271.24}  &    \textbf{-22.0} & 216.111 &   157.459 \\
M4-FG5       & no 21cm signal & 5th order      &   -1297.994   &    -33.0    & 170.291 & 157.757\\
M4-FG9       & no 21cm signal & 9th order      &   -274.688   &    -22.4   &  216.579 & 157.875 \\
\hline
\end{tabular}
\label{tab:3}
\end{table*}

The Bayesian evidence values of all models are summarized in Table~\ref{tab:3}. We compare the evidence of 44 models\footnote{
The Bayesian evidence indicates how well the model describes the data. The result of 51<$\nu$<99 MHz should not be directly compared with the result of 51<$\nu$<87 MHz.
}. The evidence values become significantly high if we assume an enhanced noise model (i.e. $A_{\rm n}$=2.25, $T_{\rm wn}$=0.015). This indicates the presence of additional noise sources such as calibration error and polarized foregrounds as discussed in \cite{2020MNRAS.492...22S} and/or the presence of unknown systematic error which is not properly modeled with the sinusoidal function. It should be mentioned that the values assumed for the enhanced noise ($A_{\rm n}$=2.25, $T_{\rm wn}$=0.015) are not physically motivated ones. Thus, additional systematic spectrum features may improve the evidence values without the enhancement of noise. We also find that the models with 5th-order polynomials have low evidence values except for HCutBW-EN. Thus the polynomial function of higher than 6th-orders is required to remove the foregrounds. 

In many cases except for FullBW-TN, the evidence is high even without the 21cm signal. Thus, we could not confirm the presence of a 21cm absorption line. In FullBW-TN, we find the statistically significant difference of $\ln Z$ between the FullBW-TN-M4-FG9 (without 21cm signal) and FullBW-TN-M3-FG9 (with 21cm signal). However, the evidence is much lower than that of FullBW-EN. This might caution that the analysis of a 21cm global signal can lead to noise-biased results. However, it should be emphasized that our results do not draw any conclusion regarding the strong absorption reported in \cite{2018Natur.555...67B} since our emulator cannot produce the flattened Gaussian shape and absorption of 500~mK reported by EDGES.

\subsection{Example: FullBW-EN-M1-FG7}

\begin{figure*}
\centering
\resizebox{180mm}{!}{
\includegraphics[width=4cm]{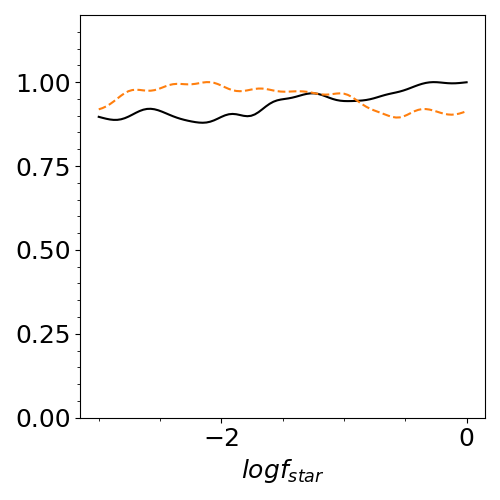}
\includegraphics[width=4cm]{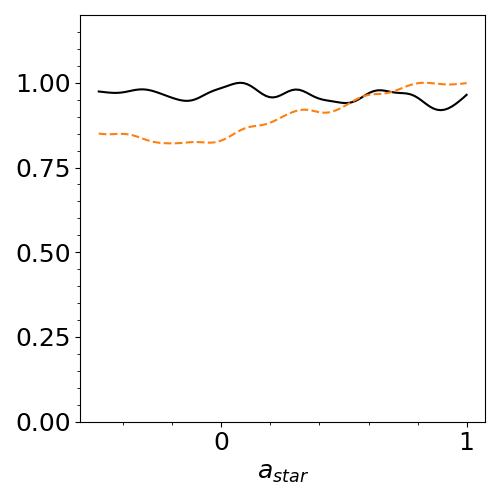}
\includegraphics[width=4cm]{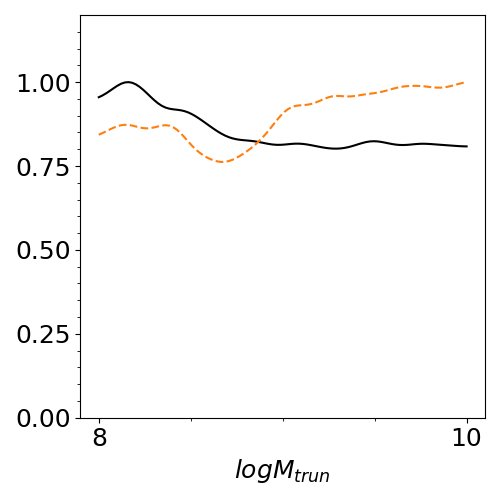}
\includegraphics[width=4cm]{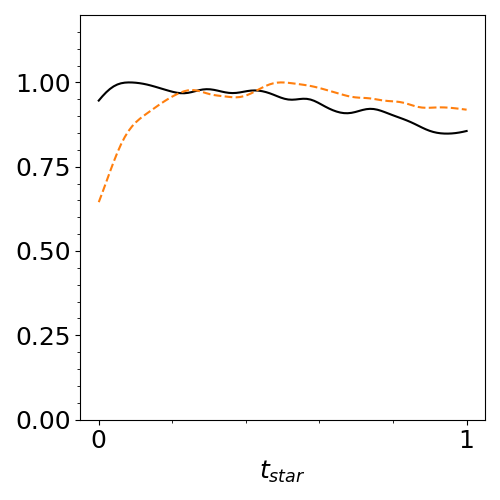}
\includegraphics[width=4cm]{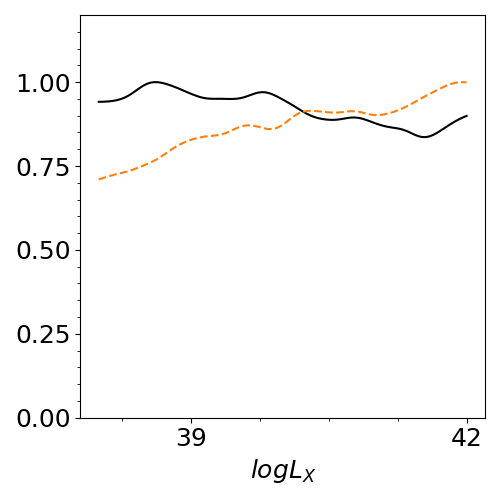}
\includegraphics[width=4cm]{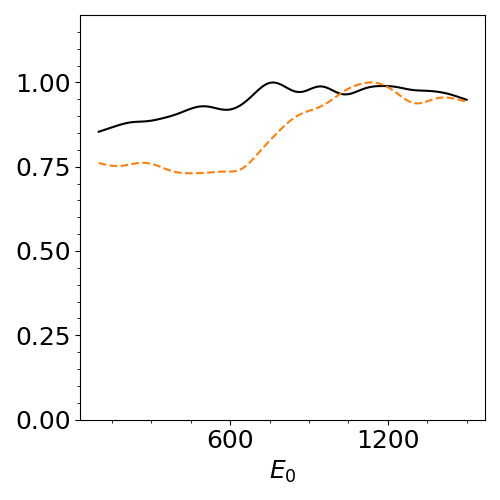}
}
\caption{Posterior distribution of astrophysical parameters in FullBW-EN-M1-FG7 (black, using 51-99 MHz).  We use our emulator of 21cm global signal, 7th-order polynomial function for the foreground, enhanced noise, and sinusoidal systematics in the analysis. Corresponding posterior of the 21cm global signal and the UVLF are respectively shown in Figure~\ref{fig:posteriorcase3gs} and Figure~\ref{fig:posteriorcase3lf}. We also show the posterior distribution of astrophysical parameters of the HcutBW-EN-M1-FG7  (orange, using 51-87 MHz) to show the influence of the frequency band. }
\label{fig:posteriorcase3}
\end{figure*}

\begin{figure}
\centering
\includegraphics[width=8cm]{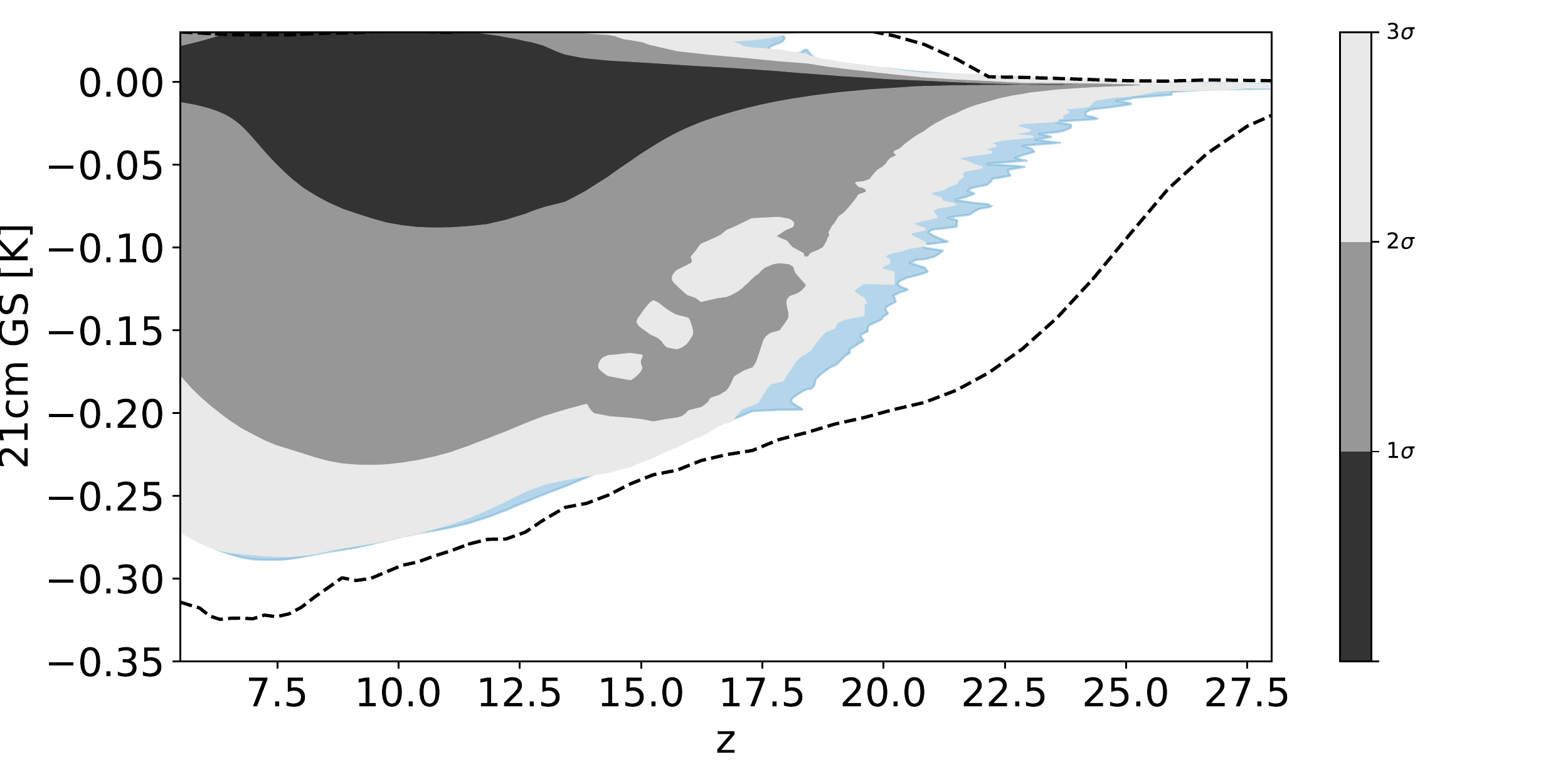}
\caption{The posterior of 21cm global signal corresponds to the parameter constraints in figure~\ref{fig:posteriorcase3}. This is the result of FullBW-EN-M1-FG7. 
The dashed lines are the min/max range of the global signal calculated using 20000 random samples within the prior in FullBW-EN-M1. The blue region corresponds to the prior samples within 3 $\sigma$. 
}
\label{fig:posteriorcase3gs}
\end{figure}

We discuss the result of FullBW-EN-M1-FG7 in detail as an example model. Figure \ref{fig:posteriorcase3} shows the posterior distribution of astrophysical parameters. The shape of the posterior distribution of all astrophysical parameters is almost flat over the prior range. Figure~\ref{fig:posteriorcase3gs} shows the 21cm global signal calculated from the posterior samples\footnote{
We use fgivenx \citep{fgivenx}.
}. The constraint is consistent with zero within 1 $\sigma$ error. The blue region shows the posterior of the 21cm global signal from random samples within priors\footnote{
Specifically, we create 20000 samples of 21cm global signal from random parameters within the prior range. The posterior density distribution, shown as the blue region, is evaluated as 3 $\sigma$ region of the random samples using fgivenx.
}. 
The figure indicates that we cannot argue the detection of the 21cm line from this model. 

The strong upper limits on the 21cm power spectrum observation measured by HERA \citep{2023ApJ...945..124H} have put the constraints on the spin temperature at $z=10.4$ and $z=7.9$ using various 21cm models including 21cmFAST. Thus, joint analysis with the power spectrum result would give a tighter lower limit on the 21cm global signal at $z<10$. In \cite{2023arXiv230103298B}, they have performed a joint Bayesian analysis using the SARAS3 and HERA observation data. Our constraints on the 21cm signal are consistent with their result while they have not used the 21cmFAST to obtain the 21cm signal and imposed a wider prior range in the likelihood analysis.

\begin{figure}
\centering
\resizebox{85mm}{!}{
\includegraphics[width=5cm]{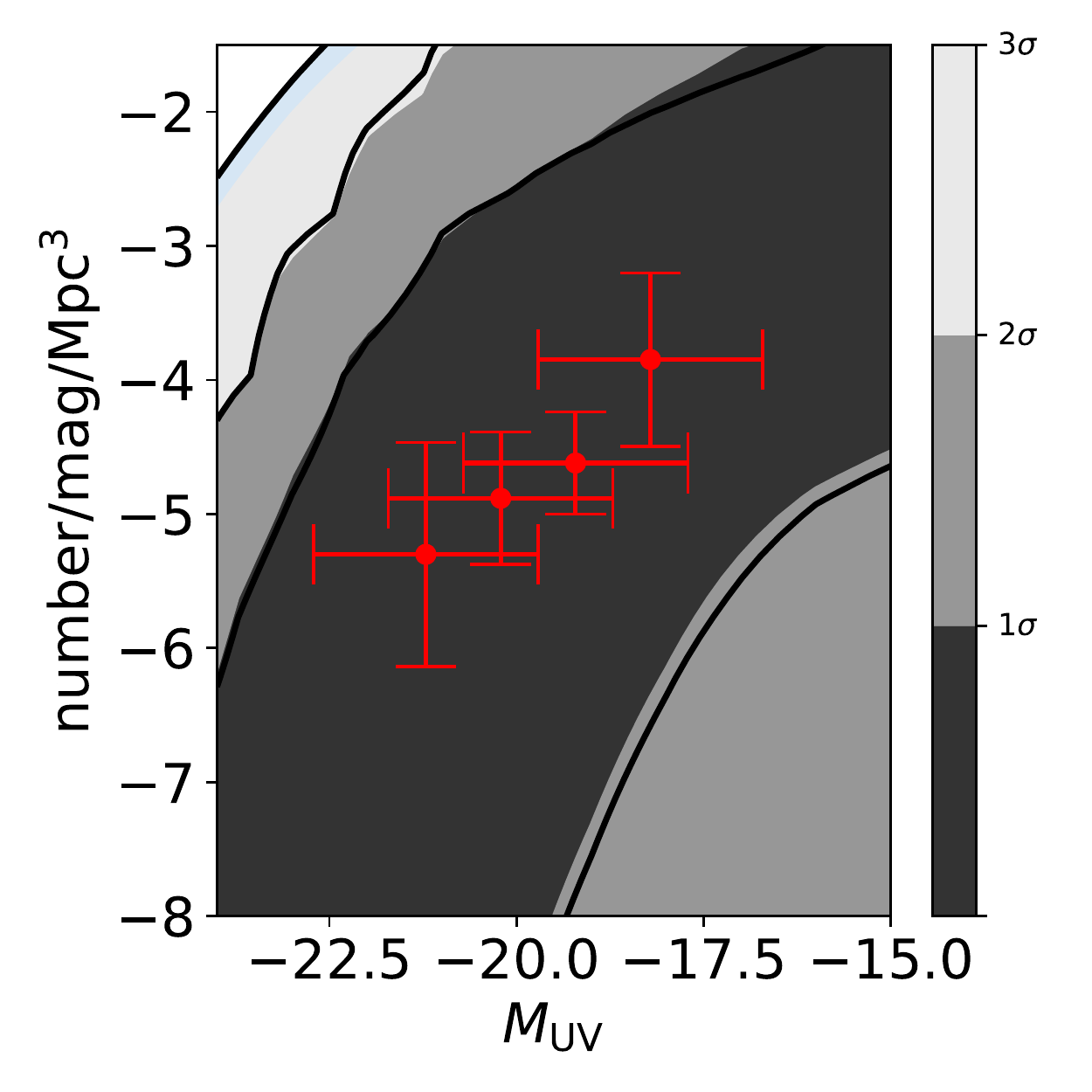}
\includegraphics[width=5cm]{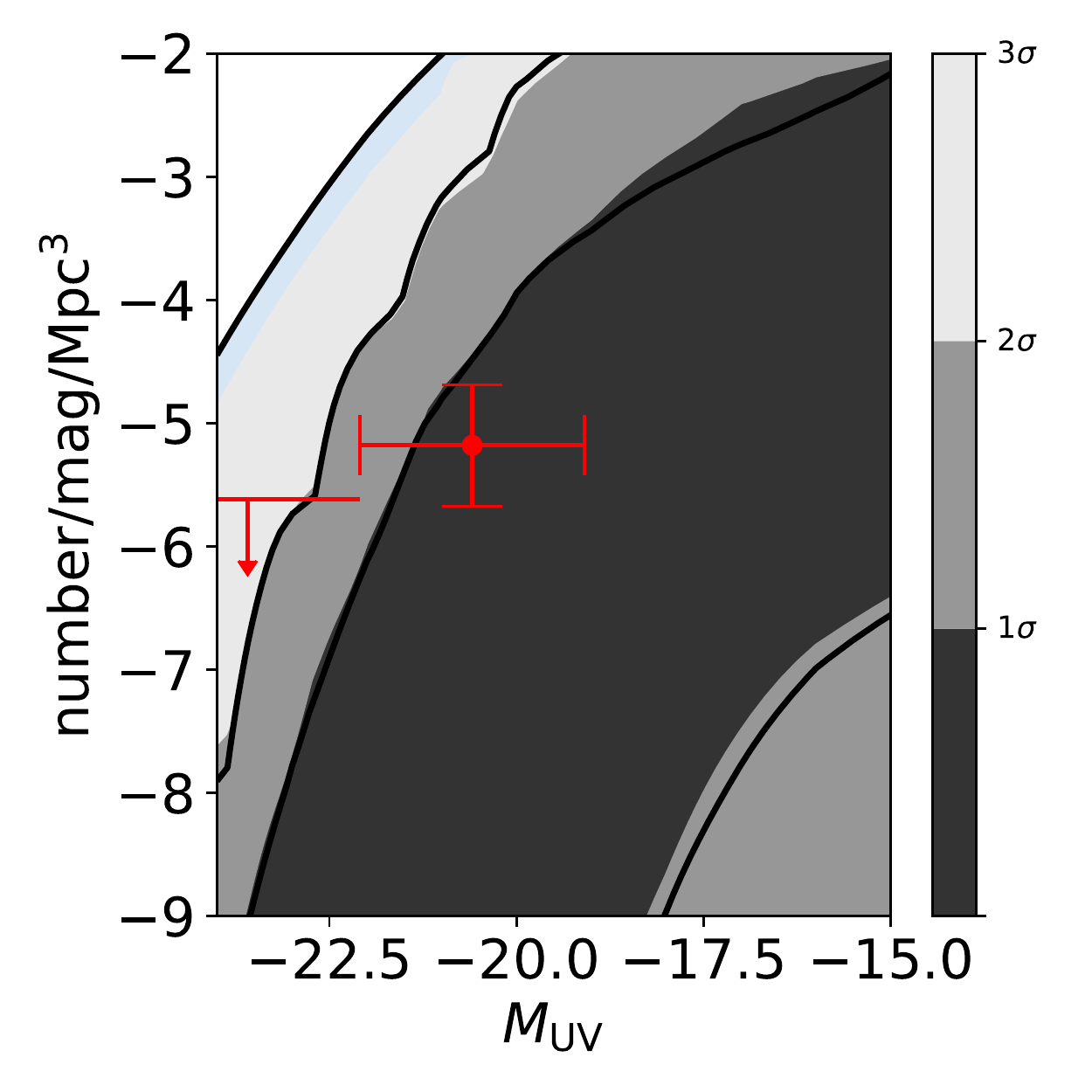}
}
\caption{The posterior of UVLF in FullBW-EN-M1-FG7. The thick line indicates the posterior of the UVLF evaluated from random  samples. Red dots, error bars, and upper limits are the observed UVLF reported \citep{Harikane2023}. }
\label{fig:posteriorcase3lf}
\end{figure}

The model of star formation rate used in the 21cmFAST can be applied to evaluate the UVLF following \cite{Park2019}. Thus, we can also obtain the posterior distribution of the UVLF.  Figure~\ref{fig:posteriorcase3lf} 
shows the constraint on the UVLF at $z=12$ and $z=16$. The results are well consistent with the recent JWST results \citep{Harikane2023}. The central value of UVLF at $z=16$ slightly offsets the centre of our confidential region.

\subsection{Comparison with other cases}

\begin{figure}
\centering
\includegraphics[width=8cm]{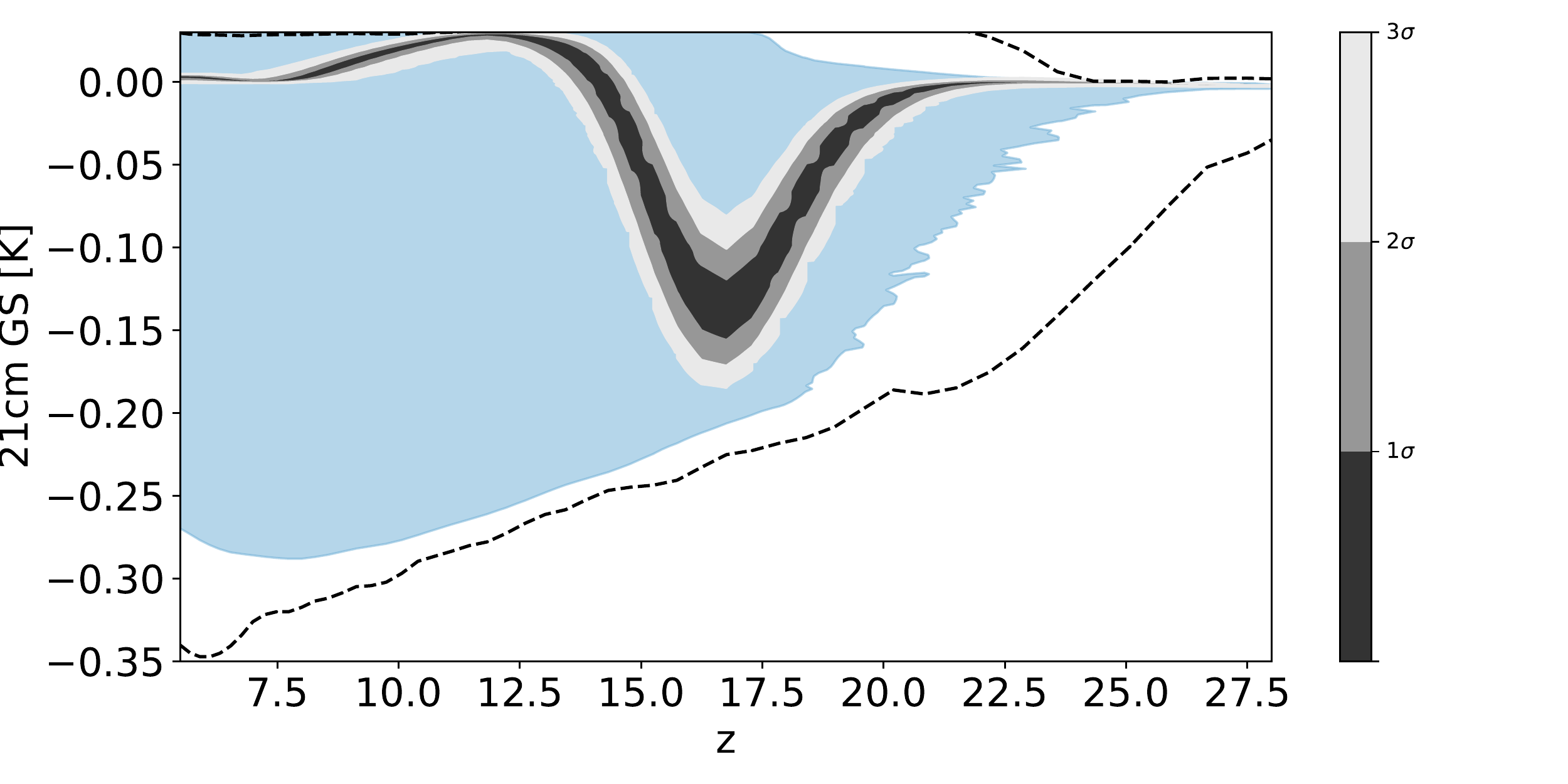}
\caption{Same as figure \ref{fig:posteriorcase3gs}, bur for the FullBW-EN-M1-FG5. The posterior of global signal (the black contour) indicates the absorption line at $z \approx 17$, however, this model has a lower evidence value as $\ln {\cal Z}=199.679$ than the other foreground models (FullBW-EN-M1-FG6 to FG9 in table \ref{tab:3}).
}
\label{fig:CASE3-ID1}
\end{figure}

We first examine the effect of the order of polynomials by comparing M1-FG5 to M1-FG9 of FullBW-EN. The constrained global signal is similar to each other and consistent with zero within 1 $\sigma$ error except the FullBW-EN-M1-FG5. The FullBW-EN-M1-FG5 with the lower polynomial order of the foreground model suggests an absorption at $z\approx 17$ with a peak amplitude of roughly 150~mK but has lower $\ln \mathcal{Z}$ (Figure~\ref{fig:CASE3-ID1}). This result caution that an improper assumption on the order of the polynomial can potentially lead to a biased detection. 

Comparing FullBW-EN-M1-FG7 and HCutBW-EN-M1-FG7, we find the constraints on the global signal are similar and consistent with zero at 1 $\sigma$. HCutBW-EN-M1-FG7 (i.e. removing data at $\nu>87 \rm MHz$) also shows weak constraints on astrophysical parameters  (Figure~\ref{fig:posteriorcase3}).

\begin{figure}
\centering
\includegraphics[width=8cm]{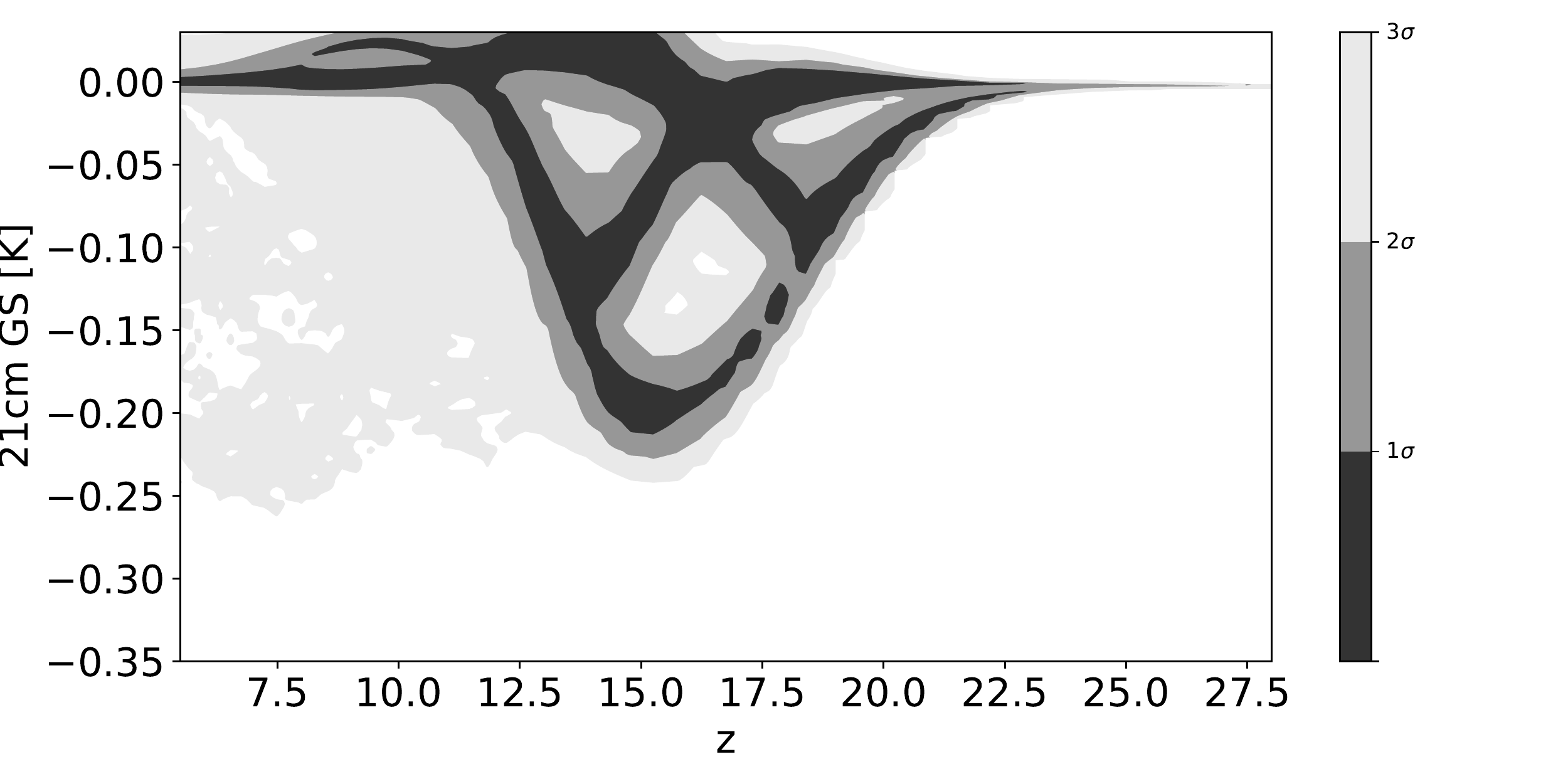}
\caption{The posterior of 21cm global signal in FullBW-TN-M1-FG7. In this model, our emulator, 7-th order polynomial function, theoretical noise and all frequency bands are used. The $f_{\rm esc}$ and primordial power spectrum are fixed. We assume the theoretical noise model. The Bayesian evidence of this model is significantly lower than our fiducial model (FullBW-EN-M1-FG7). }
\label{fig:CASE1-ID3-1}
\end{figure}

\begin{figure*}
\centering
\includegraphics[width=13cm]{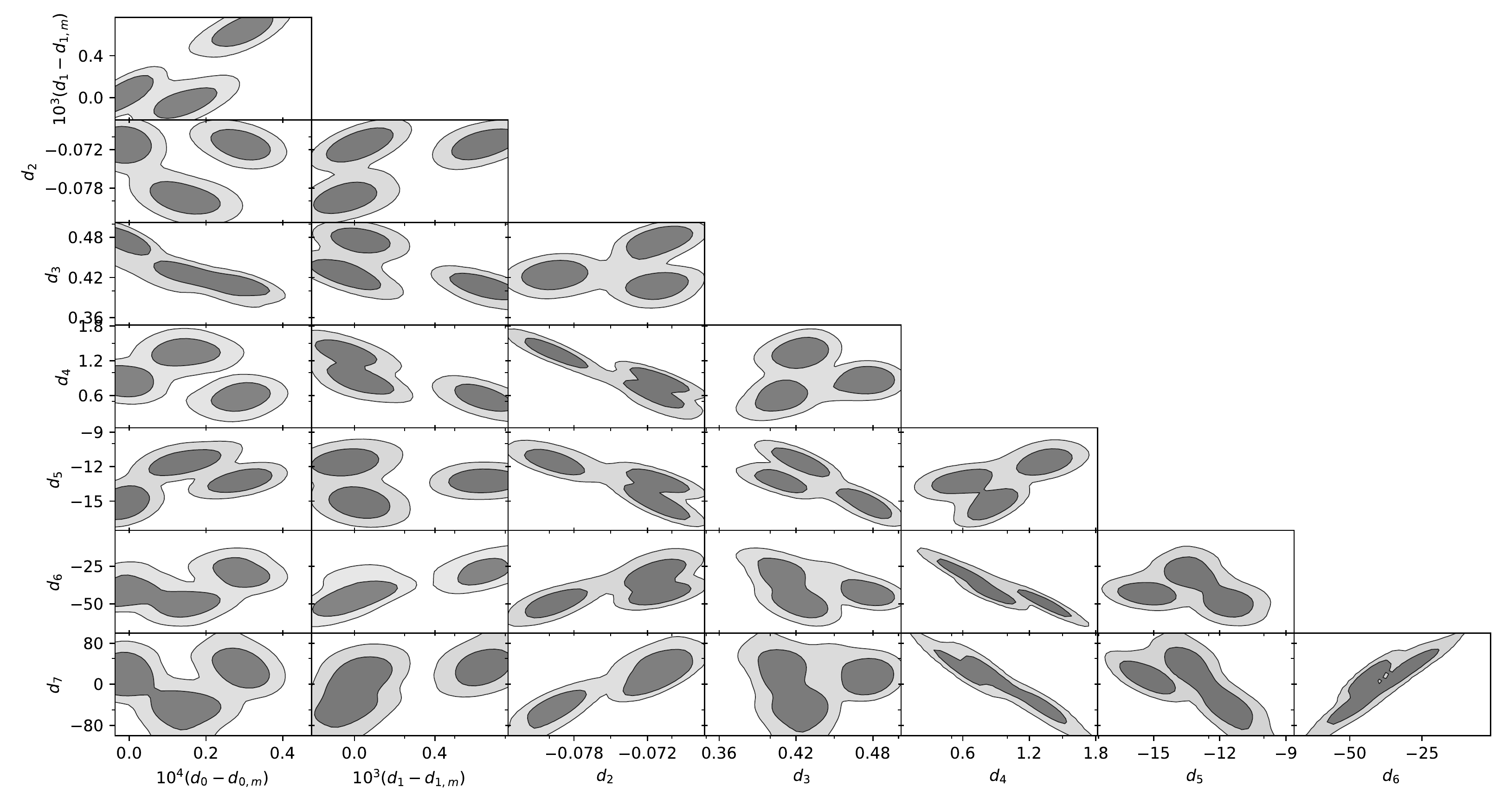}
\caption{The posterior distribution of foreground parameters and astrophysical parameters in FullBW-TN-M1-FG7. Note that, for plotting the contours of $d_0$ and $d_1$, we subtract median values ($d_{0,\rm m}=3.243$ and $d_{1,\rm m}=-2.570$) from the samples and 
multiply $10^4$ and $10^3$, respectively.
}
\label{fig:CASE1-ID3-2}
\end{figure*}

Now, we discuss the case of the other noise model, which is theoretically motivated thermal noise.
The FullBW-TN (HCutBW-TN) results are consistent with zero within 2 (1) $\sigma$ although we can find complicated structures in the posterior global signal for some cases (e.g. absorption features in Figure~\ref{fig:CASE1-ID3-1}.). The structure should be caused by clusters of posterior distributions of foreground parameters. For example, the posterior distributions of $d_0$ and $d_1$ show clusters 
as shown in Figure~\ref{fig:CASE1-ID3-2}. Even with a small difference in foreground parameters, the impact can be larger than the 21cm signal because the foreground is significantly bright. The models with theoretical noise model tend to show such clustering posterior distribution of the foreground parameters. This clustering posterior distribution of the foreground parameters might be interpreted as misfitting due to the underestimation of noise. There are other possible noise sources, such as the calibration error or the polarized foreground as described in section \ref{sec:3_method}. We checked that there are two or three peaks in the posterior distributions of astrophysical parameters. Each possible absorption line (Figure~\ref{fig:CASE1-ID3-1}) might correspond to each cluster (Figure~\ref{fig:CASE1-ID3-2}). 

\subsection{Effect of runnings}

\begin{figure*}
\centering
\resizebox{180mm}{!}{
\includegraphics[width=2cm]{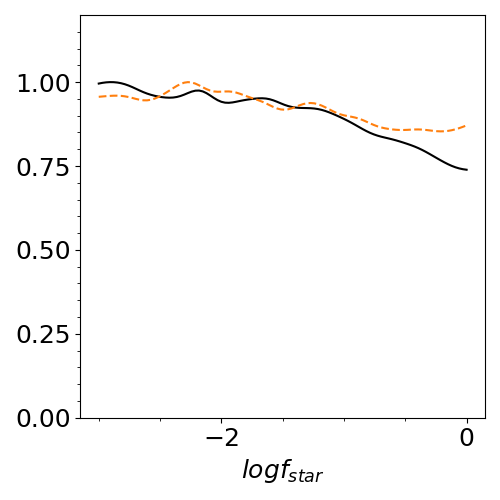}
\includegraphics[width=2cm]{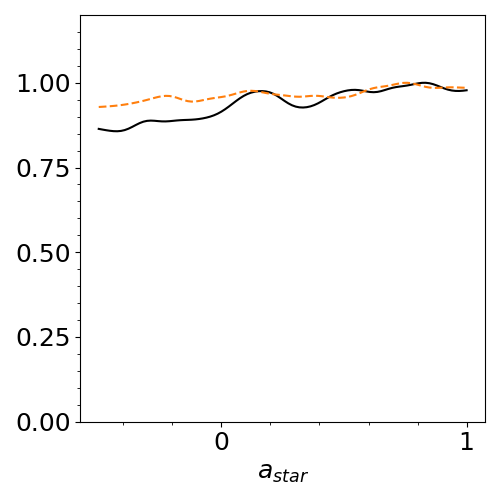}
\includegraphics[width=2cm]{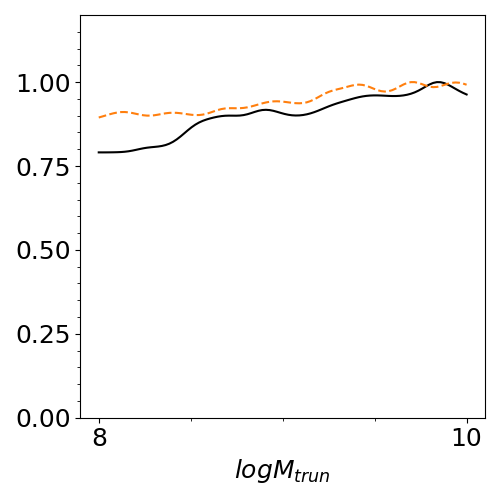}
\includegraphics[width=2cm]{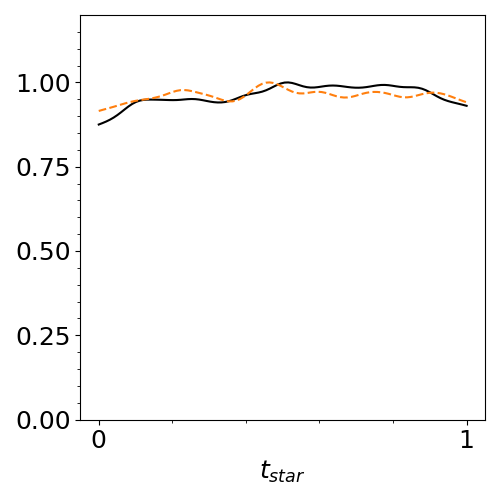}
\includegraphics[width=2cm]{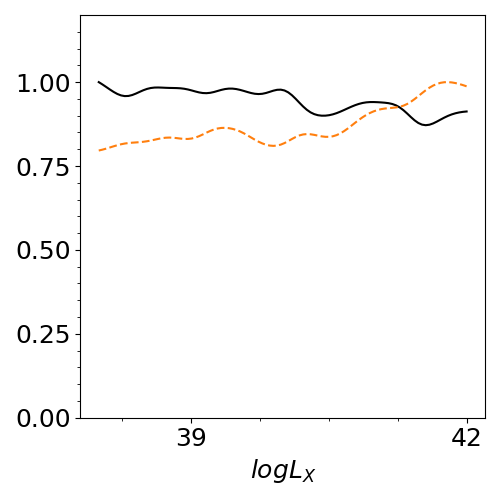}
\includegraphics[width=2cm]{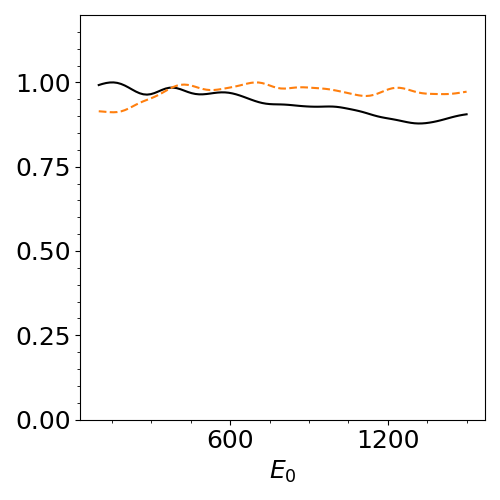}
}
\caption{Comparison of the posterior distribution of astrophysical parameters of FullBW-EN-M1-FG9 (black) and FullBW-EN-M2-FG9 (orange) to show the influence of the primordial power spectrum on the parameter constraints. }
\label{fig:CASE3-ID5-7}
\end{figure*}

\begin{figure}
\centering
\resizebox{85mm}{!}{
\includegraphics[width=5cm]{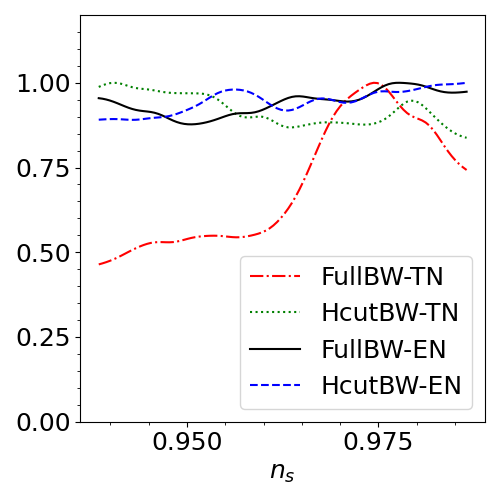}
\includegraphics[width=5cm]{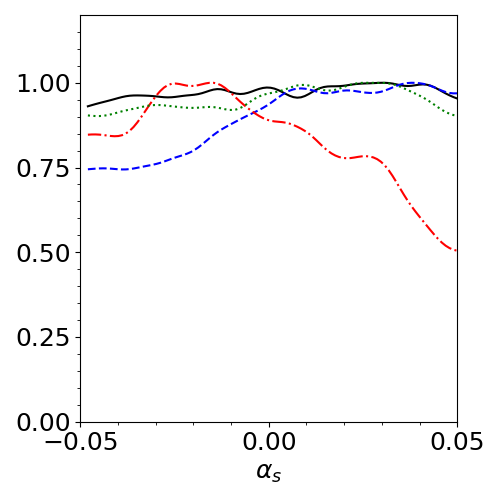}
\includegraphics[width=5cm]{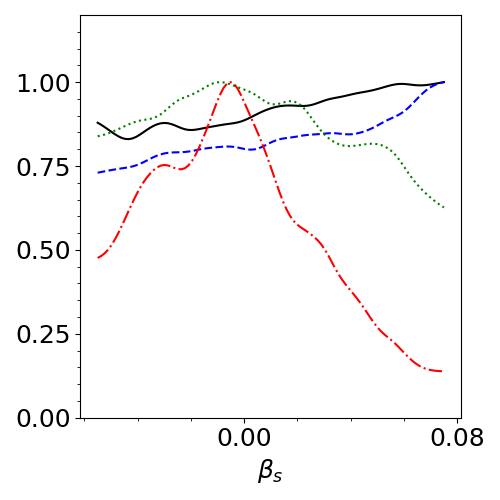}
}
\caption{The posterior distribution of primordial power spectrum parameters. We compare 4 models; FullBW-TN-M2-FG9 (red), HCutBW-TN-M2-FG9 (green), FullBW-EN-M2-FG9 (black), and HCutBW-EN-M2-FG9 (blue). }
\label{fig:nscomps}
\end{figure}

The model with higher $\alpha_{\rm s}$ and higher $\beta_{\rm s}$ can enhance the halo mass function. Therefore, the timing of Lyman-$\alpha$ coupling and X-ray heating is shifted to 
higher redshift, and hence the position of the 21cm absorption trough moves to a higher redshift as shown in Figure~\ref{fig:runningeff}. The variation of the primordial power spectrum introduces more degrees of freedom to the shape and position of the global signal profile.
Comparing the FullBW-EN-M1-FG9 (fixing $n_{\rm s},\alpha_{\rm s},\beta_{\rm s}$)
and FullBW-EN-M2-FG9 (varying $n_{\rm s},\alpha_{\rm s},\beta_{\rm s}$) can give insight into the effect of the primordial power spectrum on the astrophysical parameter constraints (Figure~\ref{fig:CASE3-ID5-7}). The constraints are weak for both cases while the posterior distributions are not perfectly consistent with each other. For the constraints on parameters of the primordial power spectrum, we also find an almost flat posterior distribution except FullBW-TN-M2-FG9  (Figure~\ref{fig:nscomps}). 
The figure indicates that constraints on the primordial power spectrum from 21cm global signal depend on the assumptions on foreground and noise in the analysis. For example, the case of FullBW-TN-M2-FG9, we find an upper limits on the $\beta_{\rm s}$, in which the constraint might be affected biased by unknown systematics and underestimated error. In any case, our analysis indicates that the 21cm global signal has the potential to constrain the running parameters if the shape of the 21cm global signal is well constrained. 

\subsection{Effect of escape fraction}

As the EDGES data corresponds to the redshift larger than 13, and the ionization would not happen at such high redshift, therefore we fixed parameters related to the escape fraction of ionizing photons in models M1 and M2. On the other hand, in the M3 model, we treat the $f_{\rm esc}$ and $a_{\rm esc}$ as free parameters. 
In the FullBW-EN-M3-FG9, we find that the posterior distributions of parameters are almost flat and the constraints on the global signal are consistent with zero.

\subsection{Luminosity function}

\begin{figure*}
\centering
\resizebox{180mm}{!}{
\includegraphics[width=8cm]{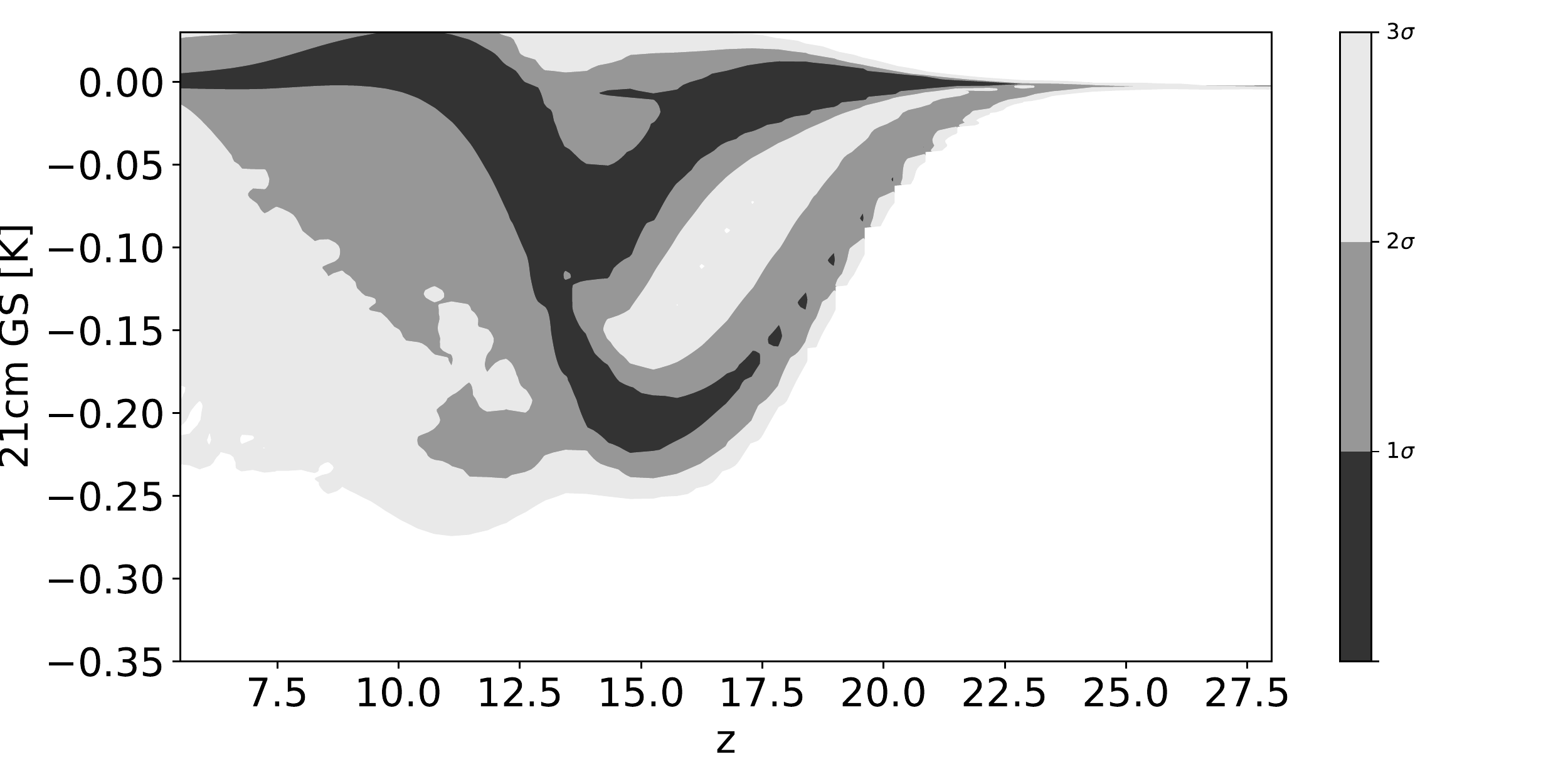}
\includegraphics[width=5cm]{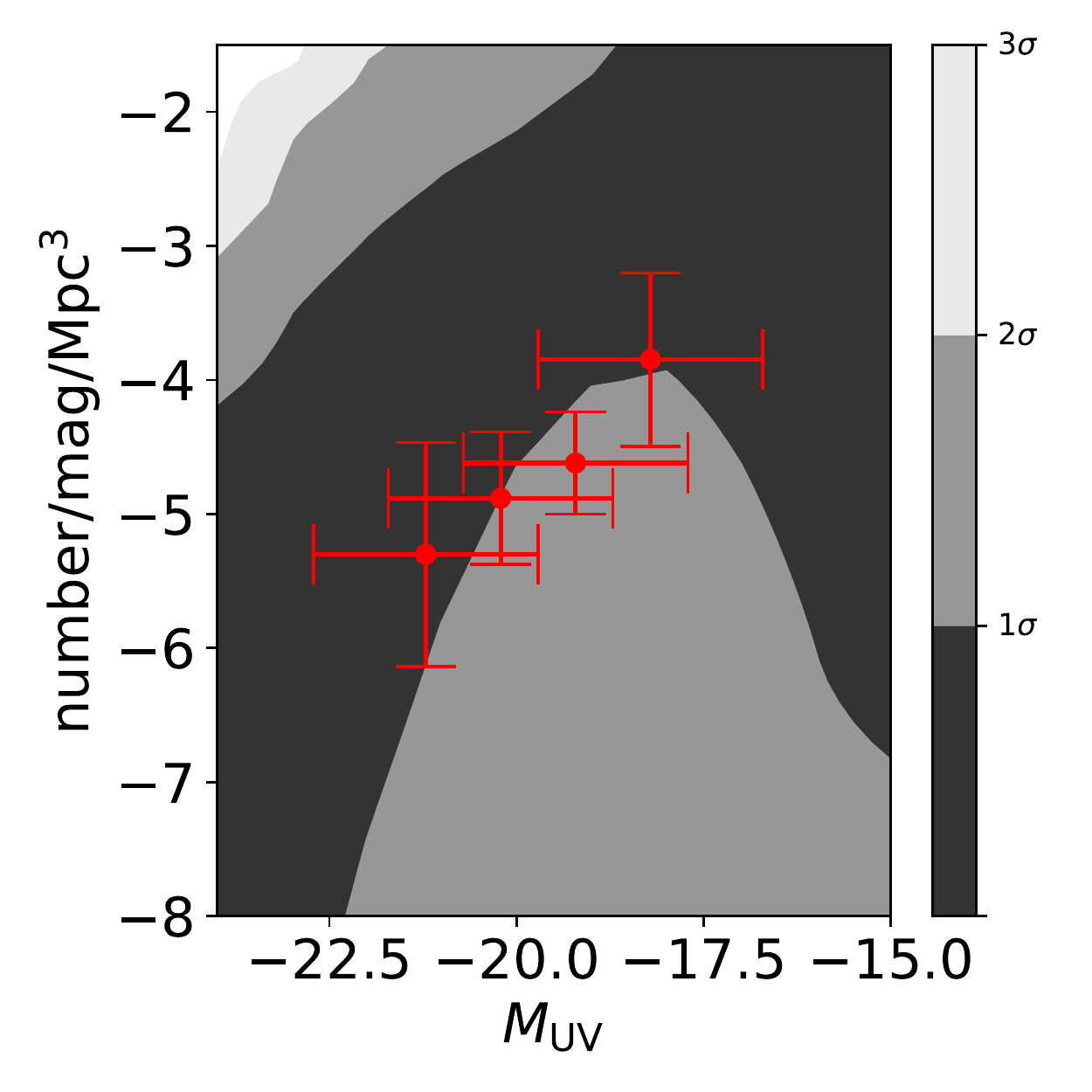}
\includegraphics[width=5cm]{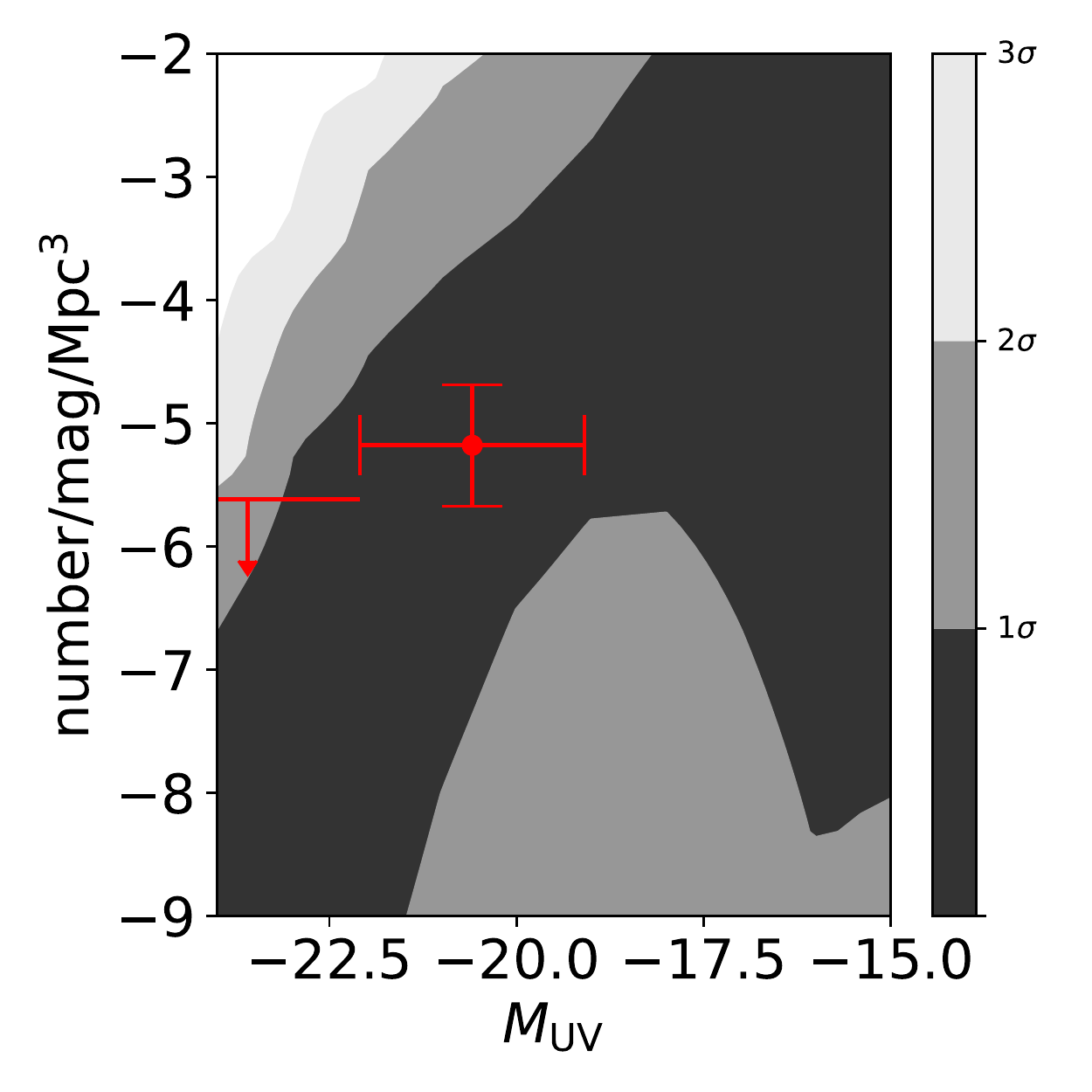}
}
\caption{The posterior distribution of 21cm global signal (left panel), UVLF of galaxies at $z=12$ (middle panel), and UVLF at $z=16$ (right panel). This result is the FullBW-TN-M1-FG9. Red points and upper limits are the results of recent JWST observation \citep{Harikane2023}. Note that the model has a lower $\ln Z$ value compared to our fiducial model. }
\label{fig:lf}
\end{figure*}

The models of high log~${\cal Z}$ provide the posterior luminosity function consistent with the JWST results as shown in figure.~\ref{fig:posteriorcase3lf}. However, the constraints on the luminosity function is weak. On the other hand, in the FullBW-TN-M1-FG9, for example, the constraints on the UVLF become tight and low values of the UVLF are disfavored as shown in Figure~\ref{fig:lf}. Furthermore, in the FullBW-TN-M1-FG9, the presence of a certain absorption line at $z>13$ is preferred in the posterior distribution of the global signal. To create the absorption feature, an effective star formation rate is required to produce a large amount of Lyman-$\alpha$ photons enough to couple the spin temperature with the cold gas temperature. Therefore the recent discovery of high UVLF at high-$z$ might be consistent with the 21cm absorption line at $z \approx 15$. It should be mentioned that the favoured parameters require a high value of $\log_{10} f_{\rm star}>-0.5$ and a low value of $t_{\rm star}<0.3$. We note that \cite{haro2023spectroscopic} recently derived the follow-up spectroscopic result of JWST and find one of the galaxies candidates at $z=16$ was the galaxy at $z=4.9$ while the galaxy suspected to be at $z>10$ was confirmed as the galaxy at $z=11.4$.

We mention that the required star formation rate to explain the EDGES absorption line has been discussed in \cite{Madau2018}. Very recently, \cite{hassan2023jwst} pointed out the UV luminosity density required to explain the EDGES absorption signal is consistent with the recent JWST observation ($z<15$) and the extrapolation of the galaxy UV observed by HST. They also show that the UVLF at $z=16$ necessary to explain the EDGES absorption is slighly higher than the UVLF extrapolated from UVLF at $z=12$. 

In \cite{2022arXiv220914312B}, they have shown such a model is consistent with current observations (e.g. neutral fraction) and pointed out the JWST can probe it. Thus, it should be interesting to perform MCMC analysis simultaneously using the UVLF, neutral fraction, 21cm global signal, and so on.

\section{Summary and future works}

We developed the ANN-based emulator to predict the 21cm global signal with 8 astrophysical parameters and 3 regarding the primordial power spectrum. The emulator can predict the 21cm global signal with the accuracy of RMSE of a few mK. Since the primordial power spectrum can control the halo mass function, we include the parameters describing the primordial power spectrum. Therefore this emulator is useful for parameter 
estimation for not only the high-$z$ star formation but also the inflation models. To demonstrate it, we performed the Bayesian analysis using the publicly available EDGES low-band data, 
from which we obtained weak constraints on the parameters. The result depends on the model of the 21cm signal, the model of the foreground, the noise level, and the frequency range used in the analysis. Using the Bayesian evidence, we compared these models. The Bayesian evidence suggests that the higher-order polynomial function is required to remove the foreground successfully and prefers the assumption of enhanced noise. The values of Bayesian evidence do not depend on the presence of the 21cm global signal and indeed the posterior distribution of the 21cm global signal is consistent with zero. 
Our analysis suggests that the detection of a 21cm global signal should be carefully understood as it depends on various assumptions on foreground, noise and so on. 

The posterior distribution of the UVLF at $z=12$ and $z=16$ derived from our Bayesian analysis using the EDGES data is well consistent with the UVLF reported by the JWST observation. 

We also found that, under our assumption of the foreground, noise, and systematics, the EDGES data cannot constrain the running parameters well. However, the 21cm signal can constrain the $\beta_{\rm s}$ if the shape and position of the 21cm signal are well constrained.

We in this work developed an emulator of the 21cm global signal including the parameters describing the primordial power spectrum. 
However we can consider not only the global signal but also other quantities such as the 21cm power spectrum, neutral fraction of hydrogen, and the evolution of spin temperature. We plan to build emulators of all such quantities and perform Bayesian analysis with current observations in future work.

\section*{Acknowledgement}

We thank Prof. J. Bowman and Dr. Monsalve for providing the weight data kindly. This work was supported in part by JSPS KAKENHI Grant Nos. 21J00416~(SY), 22KJ3092~(SY), 19K03874~(TT) and MEXT KAKENHI 23H04515~(TT). SY is supported by JSPS Research Fellowships for Young Scientists. TM is supported by JSPS Overseas Research Fellowship.

We used softwares Polychord \citep{2015MNRAS.453.4384H, 2015MNRAS.450L..61H}, fgivenx \citep{fgivenx}, 21cmFAST \citep{2007ApJ...669..663M, 2011MNRAS.411..955M, Park2019}.
Numerical computations were in part carried out on PC cluster at Center for Computational Astrophysics, National Astronomical Observatory of Japan.

\section*{Data Availability}

The EDGES data is publicly available at \url{https://loco.lab.asu.edu/edges/edges-data-release/}. We mainly use open-source software. The data used to generate the results reported in this paper are not publicly available but will be shared, on reasonable request to the corresponding author.
 



\bibliographystyle{mnras}
\bibliography{eorSY,global}








\bsp	
\label{lastpage}
\end{document}